# Data-driven Design–Test–Make–Analyze Paradigm for Inorganic Crystals: Ultrafast Synthesis of Ternary Oxides

Haiwen Dai[1], Matthew J. McDermott[2], Andy Paul Chen[1], Jose Recatala-Gomez[1], Wei Nong[1,3], Ruiming Zhu[1,4], Maung Thway[1], Samuel Morris[1], Christian Schürmann[5], Shreyas Dinesh Pethe[1], Chenguang Zhang[1], Wuan Geok Saw[6], Bich Ngoc Tran[6], Pritish Mishra[1,7,8], Fengxia Wei[4], Albertus Denny Handoko[4], Sabrine Hachmioune[4,9], Haipei Shao[10], Ming Lin[4], Chong Wai Liew[6], Kristin A. Persson[11], and Kedar Hippalgaonkar[1,4,*]

[1] *School of Material Science and Engineering, 50 Nanyang Avenue, Nanyang Technological University, Singapore 639798, Republic of Singapore.*

[2] *Materials Sciences Division, Lawrence Berkeley National Laboratory, Berkeley, California 94720, USA.*

[3] *Institute of High-Performance Computing, Agency for Science Technology and Research (A*STAR), 1 Fusionopolis Way, Singapore 138632, Republic of Singapore.*

[4] *Institute of Materials Research and Engineering, Agency for Science Technology and Research (A*STAR), 2 Fusionopolis Way, Singapore 138634, Republic of Singapore.*

[5] *Rigaku Europe SE, Hugenottenallee 167, 63263 Neu-Isenburg, Germany.*

[6] *Institute of Structural Biology, 59 Nanyang Drive, Nanyang Technological University, Singapore 636921, Republic of Singapore.*

[7] *School of Physical and Mathematical Sciences, 21 Nanyang Link, Nanyang Technological University, Singapore 637371, Republic of Singapore.*

[8] *Energy Research Institute, Interdisciplinary Graduate Programme, Nanyang Technological University, 1 Cleantech Loop, Singapore 637141, Republic of Singapore.*

[9] *Department of Chemistry, University College London, 20 Gordon St, United Kingdom.*

[10] *Department of Chemistry, Faulty of Science, National University of Singapore, 3 Science Drive 3, Singapore 117543, Singapore.*

[11] *Molecular Foundry, Lawrence Berkeley National Laboratory, Berkeley, California 94720, USA.*

* *Corresponding author: kedar@ntu.edu.sg*

*Lead contact: Dai Haiwen, haiwen.dai@ntu.edu.sg*

**Summary**

Data-driven methodologies hold the promise of revolutionizing inorganic materials discovery, but they often face challenges due to discrepancies between theoretical predictions and experimental validation. In this work, we present an end-to-end discovery framework that leverages synthesizability, oxidation state probability, and reaction pathway calculations to guide the exploration of transition metal oxide spaces. Two previously unsynthesized target compositions, $ZnVO_3$ and $YMoO_3$, passed preliminary computational evaluation and were considered for ultrafast synthesis. Comprehensive structural and compositional analysis confirmed the successful synthesis $ZnVO_3$ in a partially disordered spinel structure, validated via Density Functional Theory (DFT). Exploration of $YMoO_3$ led to $YMoO_{3-x}$ with elemental composition close to 1:1:3; the structure was subsequently identified to be $Y_4Mo_4O_{11}$ through micro-electron diffraction (microED) analysis. Our framework effectively integrates multi-aspect physics-based filtration with in-depth characterization, demonstrating the feasibility of designing, testing, synthesizing, and analyzing (DTMA) novel material candidates, marking a significant advancement towards inorganic materials by design.

## Introduction

The quest for discovering functional inorganic materials has been a perennial ambition[1], driving various scientific and technological domains in society. With the advent of artificial intelligence and data-driven approaches, the guided prediction of new materials has become increasingly promising[2–4]. While a few success stories can be found, it is undeniable that most predicted materials remain unproven through experimentation. In recent years, several studies have sought to close the design loop of inorganic materials, from prediction to realization. However, the successful realization of predicted materials remains extremely limited[2,5–7] (Table S2 summarizes recent milestones). For instance, Claudio Zeni *et al.* predicted 8,192 stable new materials with 1 made[2]. Most predictions still lead to the making of materials that were already known[5,7,8] or competing phases[6]. This discrepancy arises partly due to an overemphasis on simplified metrics, such as synthesizability (typically represented by the likelihood of a compound is made experimentally[9]), or the thermodynamic stability (typically represented by formation energy and the energy above the hull)[10–14], which are often insufficient for guiding the fabrication of new materials. Another key factor hindering the successful realization of predictions is the lack of structure-specific synthesis planning. A predicted material has a higher likelihood of being synthesized if it is both thermodynamically stable and kinetically accessible. However, when the material is metastable and surrounded by competing phases, kinetic trapping can impede the target reaction, preventing successful synthesis.

To navigate new chemical spaces, it is essential to integrate capabilities beyond these basic metrics:

- Computational stage: Rapidly identify promising and experimentally accessible target compositions or structures.
- Experimental stage: Perform high throughput experimental screening via rational synthesis planning and rapid synthesis.
- Characterization stage: Conduct detailed composition and structural characterization, especially when pure phase or high crystallinity samples are not present.

In this context, it is crucial to consider additional metrics beyond thermodynamic stability or synthesizability to assess a material's experimental accessibility during the computational stage. Oxidation state calculation provides an additional layer of screening, ensuring charge neutrality by eliminating materials that cannot achieve proper oxidation balance.[15,16]. By ranking materials based on the likelihood of achieving a specific set of oxidation states, materials with higher likelihoods are prioritized, while theoretically possible but less likely candidates are ranked lower.

Furthermore, synthesis planning guides experimental synthesis and eventually affects the accessibility of predicted materials. Yet, due to limited understanding of crystal reaction kinetics, computational synthesis planning is still in its early stages, primarily focusing on standalone precursors or environmental conditions[17–21]. Their effectiveness in achieving experimental success remains limited[5]. Heuristically, combination reactions (e.g., $CaO + TiO_2 \rightarrow CaTiO_3$) offer a simple yet effective synthesis approach proven in oxides, chalcogenides, and intermetallics. However, the lack of selectivity can result in the formation of competing phases[22,23]. By incorporating additional elements beyond the target phase, it has been shown that enhanced thermodynamic driving forces and increased selectivity against potential competing reactions is possible[24,25]. Therefore, conducting multi-aspect screening that combines oxidation state calculations, synthesis planning, and established criteria for synthesizability or thermodynamic stability is a sensible approach to identifying promising materials for experimental synthesis.

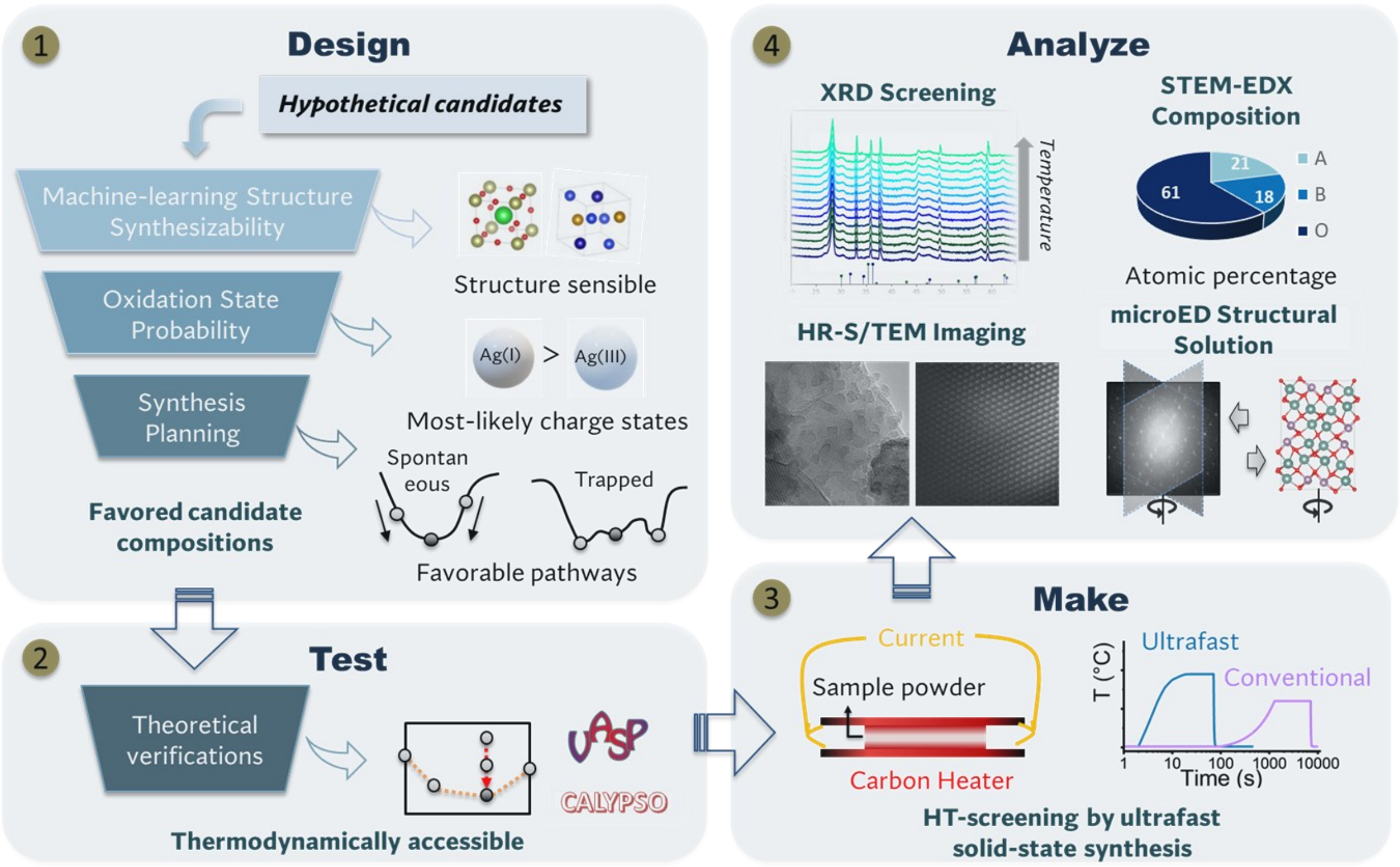

**Figure 1. Inorganic crystalline material discovery via 'Design-Test-Make-Analyze' paradigm.**

For the actual making of the material, conventional solid-state synthesis typically requires several hours or days at temperatures up to approximately 1500 °C, which creates a bottleneck for high-throughput screening and synthesis optimization[26]. Alternative techniques, such as microwave sintering[27], ultrafast sintering[28], and arc-melting[29], have demonstrated the potential to complete heating at short (typically minutes) timescales within a broad temperature range. Furthermore, automated labs have further increased the throughput[5].

In terms of characterization, structural screening can be accomplished via well-established high-throughput powder X-ray diffraction (XRD) and refinement workflows[30], complemented by more careful characterization techniques including *in situ* XRD for materials prone to instability in ambient conditions[31]. Spatially resolved electron microscopy (EM) methods provide additional structural and compositional information to validate synthetic products, especially in local regions of interest. Techniques such as S/TEM-based approaches, including elemental mapping, atomic-resolution imaging, and microcrystal electron diffraction (microED)[32], enable the measurement of chemical compositions and the identification of potentially unknown phase structures even in phase mixtures. Ambiguous phases identified by powder XRD can often be resolved through these techniques.

In this study, we aim to evaluate the feasibility of a complete material discovery framework from prediction to the discovery and synthesis of novel material. We chose to consider composition prediction and realization, a subset problem of structure prediction and realization. We explore the composition space around the well-studied $ABO_3$ system due to its diverse chemistry, crystallography, and functional attributes[33]. Different strategies involving transfer learning and machine-learned formation energy and stability predictions have been explored by the community to assess stability and predict new $ABO_3$ compounds[11,34–38]. Interestingly, most of these predicted compounds remain unsynthesized to date. Our framework (Figure 1) combines candidate design, test, fabrication, and analysis. We utilized a machine learning synthesizability model[9] and a data-driven oxidation state probability (OSP) model[16] to filter possible $ABO_3$ structures from the Materials Project[39]. The filtered candidates were then passed to synthesis planning, where energetically favorable and selective precursor sets were chosen for experimentation. Subsequent DFT optimization verified their thermodynamic stability. Since the selection of target compositions and synthesis pathways is entirely computationally driven, the possibility exists for the experimental synthesis of new compounds near the target stoichiometry, especially in cases involving compositional disorder and competing phases[6]. Hence, we focused on the A-B-O ternary system, exploring phase space close to the target stoichiometry.

Among the candidates subjected to our computational and machine learning workflow, two target phase spaces, Y-Mo-O and Zn-V-O were chosen for experimental exploration. Ultimately, $ZnVO_3$ was successfully synthesized and crystallized into a metal-deficient spinel structure confirmed through systematic thermophysical analysis and structural refinement. Efforts to synthesize $YMoO_3$ led to the formation of $YMoO_{3-x}$, with elemental composition close to 1:1:3. Our materials discovery framework, which integrates multi-aspect physics-based filtering, high-throughput screening, and localized experimental characterization, enables the identification of potentially novel phases within multi-phase mixtures. This approach thus facilitates the accelerated discovery of inorganic materials from in silico to reality.

## Results

### *Multi-aspect data-driven filtering via synthesizability, oxidation state probability, synthesis planning, and thermodynamic stability*

Figure 2A illustrates the process of machine-learning-based synthesizability and data-driven oxidation state probability filtering. DFT-computed $ABO_3$ structures were acquired from the Materials Project (v2022.10.28)[39]. A total of 255 $ABO_3$ structures were downloaded, comprising 52 tagged with experimental ICSD labels (treated as experimentally synthesized) and 203 without ICSD labels (treated as not experimentally synthesized during model training). A synthesizability model, optimized based on our previous work[9], was employed to assess the synthesizability of the 203 $ABO_3$ structures without ICSD labels (see Methods for more details). The synthesizability is quantified using a score between 0%-100%, reflecting the confidence level our model has for each compound's synthesizability. A favorable precision of 84.1% and recall of 80.2% are achieved (Figure 2B).

Recognizing that the material representation does not encompass the impact of oxidation states[40], we evaluate the likelihood of oxidation state combinations using an oxidation state probability (OSP) model based on our previous work[16]. We combine the aforementioned metrics and apply the following criteria to filter out candidates of synthesis interest:

- No polymorph exists in ICSD: Candidates with existing compositions were removed.
- Synthesizability score (SC) > 0.3: Candidates likely to be unsynthesizable were removed.
- Oxidation state probability (OSP) > 0.2: Candidates with a low likelihood of oxidation states were removed.
- No unlikely oxidation state in oxide (e.g., O in (-1) oxidation state).

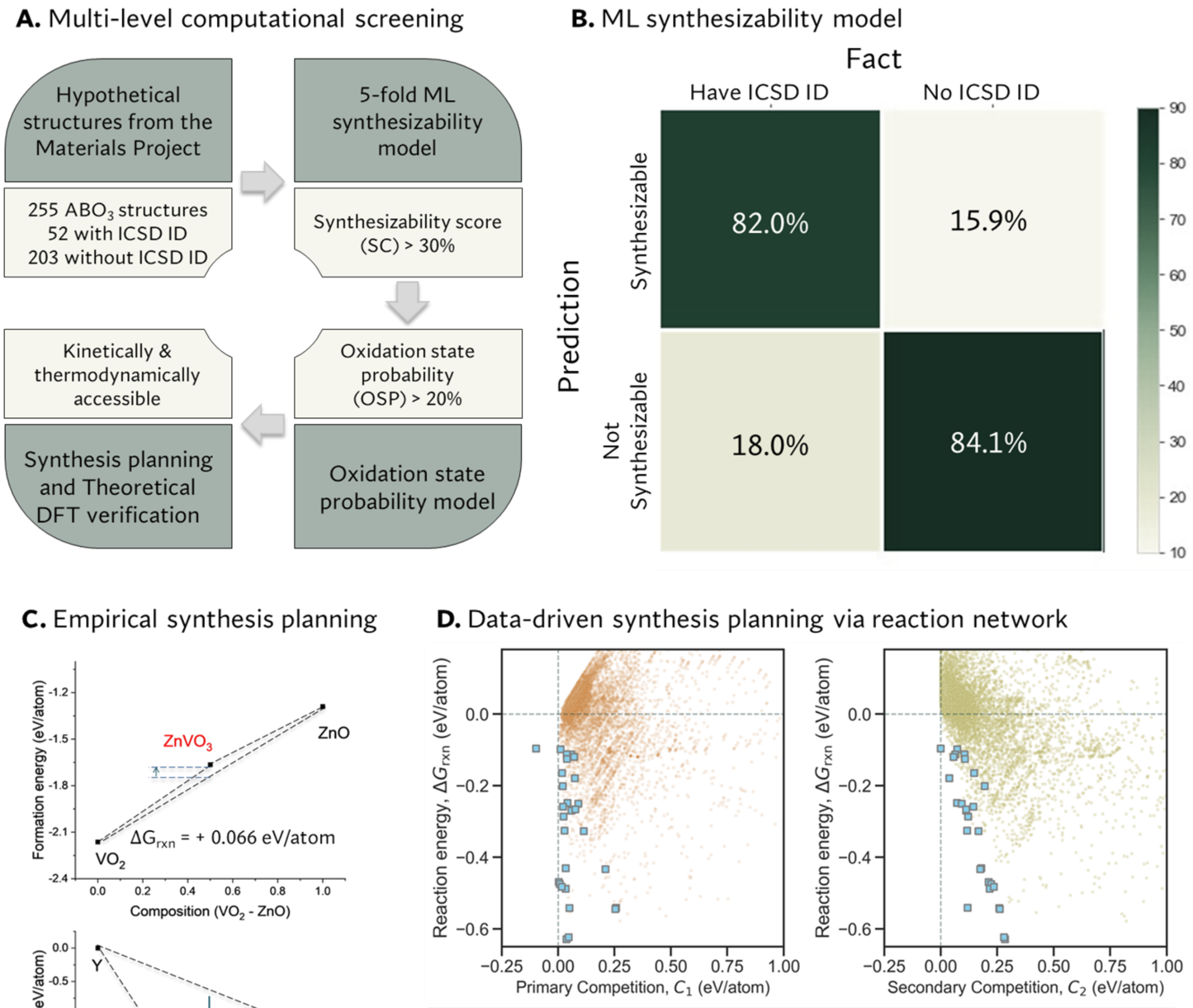

| Rank | Reaction | $\Delta G_{rxn}$ | C1 | C2 | Cost |
|---|---|---|---|---|---|
| 1 | $NaVZnO_4 + CO_2 \rightarrow ZnVO_3 + NaCO_3$ | -0.0970 | -0.0970 | 0.0000 | -0.0534 |
| 2 | $1/3\ V_3ZnO_8 + 2/3\ Mg_3V_2O_8 \rightarrow Mg_2V_2O_7 + 1/3\ ZnVO_3$ | -0.0017 | 0.0089 | 0.0073 | 0.0071 |
| 3 | $V_5O_{12} + 5\ ZnO \rightarrow ZnVO_3 + 2\ V_2Zn_2O_7$ | -0.0469 | 0.0185 | 0.0091 | 0.0077 |
| 4 | $0.2\ V_2ZnO_4 + 0.6\ Sr_5V_3O_{13} \rightarrow Sr_3V_2O_8 + 0.2\ ZnVO_3$ | -0.0407 | 0.0120 | 0.0181 | 0.0095 |
| 5 | $SrV_2ZnO_7 + K_2V_3O_8 \rightarrow ZnVO_3 + K_2SrV_4O_{12}$ | 0.0143 | 0.0191 | 0.0004 | 0.0102 |

Unit of energy: eV/atom

**Figure 2. Multi-aspect computational filtering workflow applied to ABO₃.** (A) Multi-aspect computational filtering workflow; (B) Confusion matrix of machine learning synthesizability

model; Materials with an ICSD ID were considered synthesized during training, and vice versa; (C) Empirical planning using oxide precursors towards $ZnVO_3$ and $YMoO_3$ with largest thermodynamic driving forces $\Delta G_{rxn}$ within respective ternary phase diagram; (D) Data-driven synthesis planning for $ZnVO_3$ showing the thermodynamic driving forces and selectivities of 8424 synthesis reactions. Reaction energy ($\Delta G_{rxn}$) vs. primary competition ($C_1$), vs. secondary competition ($C_2$), and 5 most selective pathways are shown. A highly selective reaction shall have a large thermodynamic driving force $\Delta G_{rxn}$, small primary competition ($C_1$) and secondary competition ($C_2$). Reactions on the three-dimensional Pareto front are plotted as blue squares.

The choice of SC and OSP minimizes the exclusion of viable candidates, acknowledging that multiple filtering stages are inherent in the selection process. By setting a lower threshold at each step, we aim to reduce the risk of omitting promising materials. At the 0.3 threshold for SC, the model achieves an 87% recall and 79% precision. This adjustment —contrasted with the confusion matrix evaluated at a 0.5 threshold (Figure 2B)— results in a 7% decrease in false negatives, with only a 3% reduction in precision. Similarly, for OSP, a threshold of 0.2 is selected, yielding a high recall of 92%. This threshold effectively filters out highly unlikely candidates, prioritizing the identification of synthesizable and charge-balanced non-ICSD materials.

**Table 1.** Selected candidates ranked by SC * OSP score. 10 non-ICSD tagged candidates passed synthesizability and charge probability filtering.

| No. | Chemical formula | SC | OSP | Most likely oxi-state combo | SC * OSP | Reported | Reported precursors |
|---|---|---|---|---|---|---|---|
| **1** | $ScFeO_3$ | 56.48% | 95.10% | [3], [3], [-2] | 53.72% | True | $Sc_2O_3 + Fe_2O_3$ |
| **2** | $CoVO_3$ | 54.30% | 90.41% | [5], [1], [-2] | 49.09% | True | $VO_2 + CoO$ |
| **3** | $NiVO_3$ | 48.13% | 73.79% | [5], [1], [-2] | 35.52% | True | $VO_2 + NiO$ |
| **4** | $CrWO_3$ | 49.33% | 60.06% | [3], [3], [-2] | 29.63% | True | $Cr + WO_3$ |
| **5** | $YMoO_3$ | 48.21% | 48.25% | [3], [3], [-2] | 23.26% | Nil | Nil |
| **6** | $YWO_3$ | 36.69% | 60.06% | [3], [3], [-2] | 22.04% | Nil | Nil |
| **7** | $ScVO_3$ | 31.84% | 62.89% | [1], [5], [-2] | 20.03% | True | $ScVO_4 + H_2$ |
| **8** | $CrVO_3$ | 33.68% | 56.15% | [3], [3], [-2] | 18.91% | True | $VCrO_4 + H_2$ |
| **9** | $ZnVO_3$ | 36.75% | 33.44% | [2], [4], [-2] | 12.29% | Nil | Nil |
| **10** | $AgRuO_3$ | 35.97% | 21.04% | [1], [5], [-2] | 7.57% | True | $Ag_2O + KRuO_4$ |
| **Filter criteria** | | >30% | >20% | N/A | N/A | Nil | N/A |

Notes: SC: Synthesizability score; OSP: oxidation state probability score; Reported precursors: Precursors reported to synthesize in the literature. Oxidation state of oxygen is fixed at -2.

Ten candidates were selected based on a combined score (SC * CP), as depicted in Table 1. Subsequently, a more comprehensive literature review identified that the 7 out of 10 candidate compositions had been experimentally reported, within which 4 can be produced via simple combination reactions. Notably, candidates without previous reports, $YMoO_3$, $YWO_3$, and $ZnVO_3$, were identified as charge-neutral and potentially synthesizable. Due to the synthesis challenges associated with the refractory nature of tungsten oxides, $YWO_3$ was excluded from further evaluation. Supplemental Note S1 provides stepwise guidance for combined SC and OSP models.

We aim to identify a simple yet effective synthetic route to the target compound by integrating empirical insights with data-driven thermodynamic analysis. Initial efforts focus on direct reactions for their practical simplicity and efficiency; more complex pathways beyond the target composition are explored when direct reactions are insufficient. We first consider combination reaction via binary oxide and metal precursors (Figure 2C), iterating through all potentially feasible combination reactions directly yield the target using stable reactants from the ternary *A*-*B*-O phase diagram (the Materials Project), and choose the reaction with the largest thermodynamic driving force. While target formation is possible with many oxide or metal precursors, competing reactions can divert reaction energy away from the desired product. For instance, the synthesis of $LiBaBO_3$[41] and $BaTiO_3$[42] from binary oxide precursors is often impeded by intermediates from undesired interfacial reactions. To avoid competing reactions, a more expansive synthesis planning considering additional elements[24] (i.e., beyond A, B, and O) was achieved via new thermodynamic selectivity metrics[42] integrated into reaction network analysis[23]. From these algorithms, pathways with the highest selectivity and thermodynamic driving forces can be selected for experimental trials (Figure 2D).

In the case of $ZnVO_3$, $ZnO + VO_2$ was determined as the most likely combination reaction precursors, despite having a positive reaction energy (Figure 2C, upper plot). Additionally, following expert advice, $0.5\ V_2O_5 + Zn + (0.5\ O_2) \rightarrow ZnVO_3$ was considered, acknowledging a slight oxygen intake. Data-driven reaction network analysis (Figure 2D) predicted that $NaVZnO_4 + CO_2 \rightarrow ZnVO_3 + NaCO_3$ was the only favorable reaction, with a combined cost score of -0.0534 eV/atom. Visualizing the reaction energy against competition reactions in the upper plots of Figure 2D, the reaction exhibits a thermodynamic driving force $\Delta G_{rxn}$ ~ -0.097 eV/atom. Primary competing reactions (denoted as $C_1$), which represent other possible reactions between the precursors, returned the same value as the target reaction, suggesting that the target reaction is the most favorable. Secondary competing reactions (denoted as $C_2$) of the predicted products were minimally favorable ($C_2 = 0$). Therefore, $NaVZnO_4 + CO_2 \rightarrow ZnVO_3 + NaCO_3$ is suggested as the most selective and spontaneous reaction among 8424 predicted synthesis reactions. More detailed explanation of the cost-ranking scheme can be found in the Methods section. However, in practical experimentation, controlling exact $CO_2$ intake proves challenging. Other reactions suggested positive cost values, indicating challenges in devising thermodynamically favorable and selective synthesis routes towards $ZnVO_3$. Consequently, binary oxide-metal combination reactions remain the most practical approach.

Similarly, for $YMoO_3$, the combination reaction of Y + $MoO_3$ was preferred owing to its substantial thermodynamic driving force of over -1.9 eV/atom (Figure 2A, lower plot). Reaction network analysis (Figure S1) identified three metathesis reactions described by $NaMoO_2$ + YO$X$ → $YMoO_3$ + Na$X$ (where $X$ = F, Cl, Br) that rank in the top five among 1278 predicted synthesis pathways. Yet, direct synthesis of $NaMoO_2$ is practically difficult, given the explosive nature of bulk sodium[43]. We approached the predicted pathway as a potential intermediate step and devised two alternate chemical reactions: (a) $Na_2MoO_4$ + 2 YO$X$ + Mo → 2 $YMoO_3$ + 2 Na$X$, considering that $Na_{0.97}MoO_2$ was synthesized via $Na_2MoO_4$ and Mo[44], and (b) 3 $Na_2CO_3$ + 2 $YX_3$ + 2 $MoO_2$ → 2 $YMoO_3$ + 3 $CO_2$ + 6 Na$X$, based on the analogous synthesis of $LiMoO_2$ via $MoO_2$ and $Li_2CO_3$[45]. The latter form is also similar to assisted metathesis routes used in the selective synthesis of $YMnO_3$ and $Y_2Mn_2O_7$[46]. The third path (Figure S1, table) is deemed infeasible due to challenges in controlling the stoichiometry of partially oxidized $Mo_9O_{26}$ reactant. The fourth path is dismissed owing to the lack of a thermodynamic driving force.

Utilizing Density Functional Theory (DFT), we assessed the energy over the convex hull ($E_{hull}$) as a metric for thermodynamic stability[47]. An $E_{hull}$ below a reasonable threshold, typically ranging between 70-100 meV/atom, has been used to predict materials' stability and to estimate experimental accessibility[10–12]. Although the threshold can be system-dependent, a large $E_{hull}$ typically suggests poor stability, which makes experimentally accessing it challenging. The compounds for convex hull construction were queried from the Materials Project, with the search limited to those calculated by the GGA+U method. Candidate compositions exhibiting large $E_{hull}$ were re-optimized via CALYPSO, searching for lower energy polymorphs. Consequently, *Pnma* $YMoO_3$ and $P2_1/c$ $ZnVO_3$ polymorphs of 30 and 54 meV/atom, respectively, indicated experimental accessibility (See Figure S2 for DFT optimization details). For comparison, the top three previously reported candidates in Table 1, $ScFeO_3$, $VNiO_3$, and $VCoO_3$, show $E_{hull}$ values of 73, 30, and 22 meV/atom, respectively. This highlights the effectiveness of our computational candidate screening mechanisms.

We note that our DFT screening was limited to fully ordered structures, potentially overlooking low-energy disordered phases. The results suggest that the target composition (at least one polymorph structure) is thermodynamically accessible. Given the computational challenges associated with enumerating all possible partially disordered configurations, we will next use high-throughput experiments and ad-hoc DFT calculations to refine the local energy landscape and identify the potential occurrence of these disordered phases. This is an important, often overlooked, step to look at competing phases in a local chemical space.

***Case Study 1: High-throughput experimental synthesis targeting the predicted composition $ZnVO_3$***

Experimental screening via high throughput ultrafast synthesis is illustrated in Figure 3A. Precursors were mixed via ball milling according to stoichiometric ratios. This process builds upon ultrafast sintering techniques[28] for the purpose of synthesizing highly crystalline inorganic powdered materials. We note that "ultrafast sintering/synthesis" refers to rapid thermal processing and is distinct from ultrafast spectroscopy or microscopy, which operate below microsecond timescales. In a typical experiment, powder mixtures were enveloped within a <1 mm thick graphite enclosure, acting as a miniscule reaction chamber. It is crucial for the precursor powders to be fully wrapped by thin graphite paper to mitigate potential phase inhomogeneity issues caused by slow heat conduction. The enclosure is then placed between a pair of carbon-based filaments capable of Joule heating to attain the intended temperature. Owing to the high thermal conductivity of carbon and Joule-based heating, heat generation and dissipation are much faster than typical solid-state synthesis. Figure 3B shows a photo series of the heating stage during a 15s heating cycle. Leveraging the highly thermally conductive carbon-based chamber, a heating and cooling rate of $\sim 10^3$ K/s can be achieved, as shown in the measured time-temperature profiles in Figure 3C and 3D. Rapid heating and cooling minimize the time at temperatures outside target ranges, helping to reduce unwanted product decomposition and/or precursor evaporation[28], and leading to full rapid synthesis completion of high quality crystalline materials within just two minutes (further information provided in Methods). Pt foil enclosure may be used for candidates with high oxidation states.

The synthesis optimization process for $ZnVO_3$ begins with two precursor sets: (a) $VO_2$ + ZnO and (b) $V_2O_5$ + Zn. Analyzing the XRD patterns in Figure 3E, we observe the partial reaction between the precursors $VO_2$ and ZnO from 300 °C, resulting in the formation of different Zinc Vanadium Oxides with spinel structures. As the temperature increases, a complete transformation occurs, yielding a phase preliminarily indexed to spinel Zn-V-O family, whose difference with existing structures will be discussed later. Upon reaching higher temperatures of 1100 °C and beyond, minor ZnO appears in addition to the spinel phase. Notably, typical spinel Zn-V-O (e.g. $ZnV_2O_4$ or $Zn_3V_3O_8$, metal to oxygen ratio 3:4) contains fewer oxygen atoms than $ZnVO_3$ (metal to oxygen ratio 2:3), indicating oxygen deficiency. In an effort to elevate oxygen intake, the reaction environment was transitioned from argon to ambient air in a furnace. The oxidation resulted in the formation of $Zn_2V_2O_7$ confirmed by XRD analysis and Rietveld refinement (Figure S3A and S3C). Intriguingly, $Zn_2V_2O_7$ possesses one additional oxygen atom compared to $ZnVO_3$ ($Zn_2V_2O_7$ vs. $Zn_2V_2O_6$), indicating greater oxygen incorporation. *In situ* XRD was carried out to understand whether there was a formation of high-temperature metastable phases that decompose at room temperature (Figure S3E). During the high temperature stages, no additional phase change was observed, suggesting that the spinel phase is stable.

Guided by heuristic insights, an additional synthesis pathway was considered: 0.5 $V_2O_5$ + Zn + (0.5 $O_2$) → $ZnVO_3$. From the XRD analysis in Figure 3F, the precursor mixture of $V_2O_5$ and Zn exhibited complete conversion into low-crystallinity spinel Zinc Vanadium Oxide following ball milling, as indicated by Rietveld refinement in Figure S3D. Commencing at 500°C, minor precipitation of ZnO becomes evident, which is subsequently incorporated back into the spinel lattice at 1100 °C. Additional *in situ* XRD, illustrated in Figure S3F up to 900 °C, confirms the prevalence of the spinel phase across all temperature ranges. This observation aligns with the results obtained using the $VO_2$ + ZnO precursors. On the other hand, under oxidizing conditions, the decomposition and oxidation of spinel results in the emergence of a mixture containing five distinct phases (Figure S3B, blue pattern), including two phases featuring the fully oxidized V(5+) state, $Zn_2V_2O_7$ and $Zn_3(VO_4)_2$.

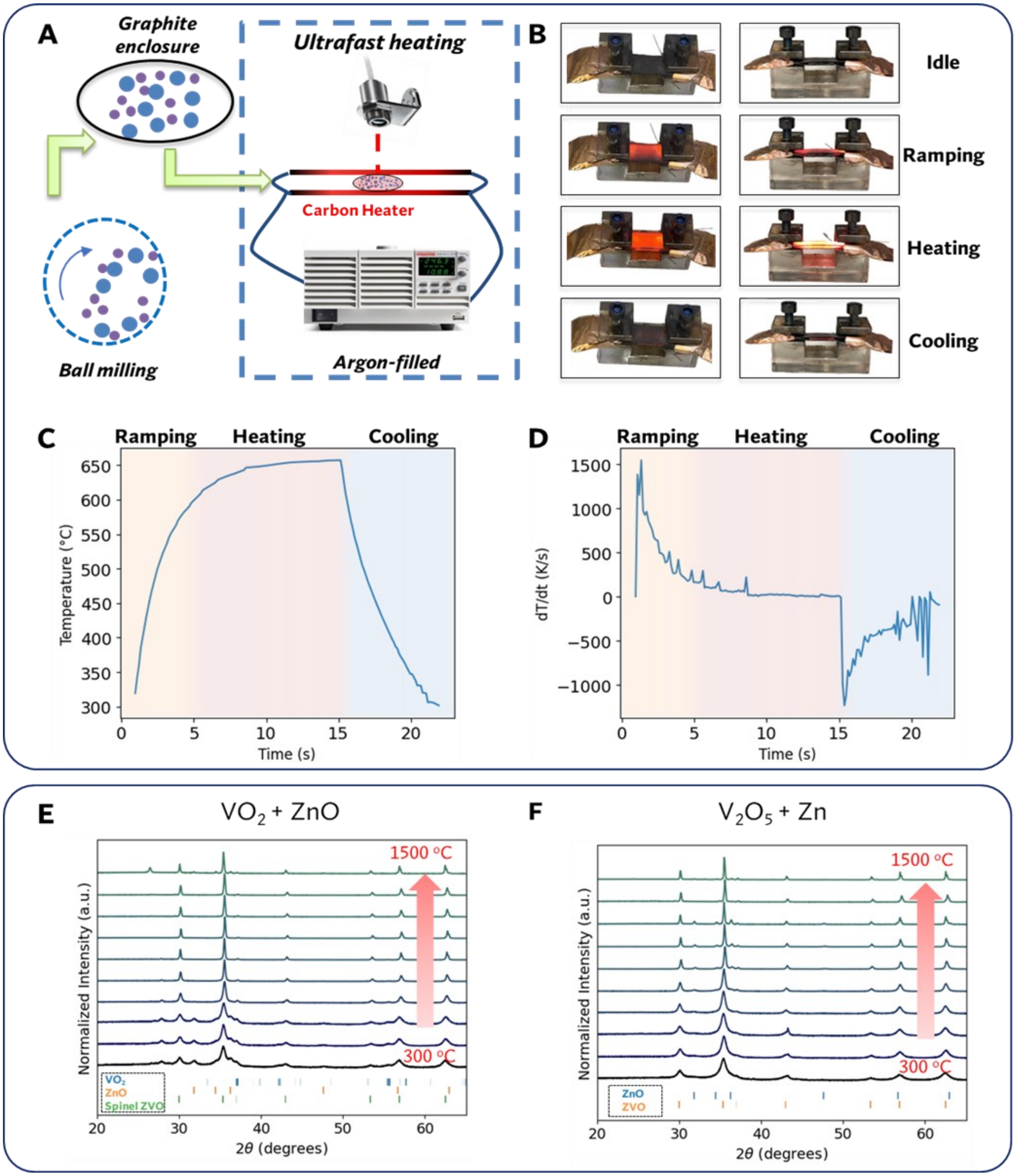

**Figure 3. Ultrafast synthesis and XRD analysis targeting the predicted composition $ZnVO_3$.** (A) Illustration of high throughput ultrafast synthesis setup; (B) photo series during a 15s heating cycle; with corresponding (C) time-temperature profile; and (D) temperature derivative-time profile. Stacked XRD spectrum of synthesized powders via precursor set (E) '$VO_2$ + ZnO' and (F) '$V_2O_5$ + Zn', showing temperature series of XRD patterns. Temperature for actual synthesis is calibrated at 15s of constant current supply. Heating and cooling rate of ~ $10^3$ K/s is measured as the largest temperature derivative during ramping and cooling. Spinel Zinc Vanadium Oxide (ZnVO) phase(s) are predominant in all synthesis towards $ZnVO_3$.

Spinel Zn-V-O compounds share similar XRD characteristics, consisting of ordered and partially disordered phases materials with the same framework[48], including but not limited to $ZnV_2O_4$, $Zn_2VO_4$ and $Zn_3V_3O_8$: this further complicates structural assignment. We then conducted a series of additional experimental characterization tests to understand the structure and composition attributes of the spinel powders. Figure 4A displays the TGA plot in an inert atmosphere from room temperature to 900°C, the maximum temperature our furnace can achieve, using ball-milled $VO_2$ and ZnO as reactants (which produced a pure spinel phase). The weight change during TGA is closely related to the solid oxygen content in the product. A minor weight fluctuation within (-0.5%, 1.5%) is observed throughout the process, with a slight 0.6% increase in solid oxygen at the final temperature of 900°C. The initial weight increase may be attributed to non-stoichiometric fluctuations in $VO_2$, which exists in several non-stoichiometric forms (Y.V. Kuznetsova et al., J. Phys.: Conf. Ser., 2019). Further approaching 100% is related to the depletion of $VO_2$ and reaction towards stoichiometric $ZnVO_3$. This solid oxygen level (0.6%) closely matches that of $ZnVO_3$, being 4.4% different from $Zn_2V_2O_7$ and 3.6% different from $Zn_3V_3O_8$. This suggests that the actual stoichiometry is much closer to $ZnVO_3$ rather than fully occupied spinel Zn-V-O structures. Conversely, in air, a 4.5% weight gain in solid oxygen triggers the transformation into $Zn_2V_2O_7$, which is consistent with the experimentally indexed $Zn_2V_2O_7$ phase (see Table S4 for details of TGA). Figure 4B shows the EDX spectrum with quantification. Multiple regions across the sample were analyzed, confirming an elemental ratio of Zn : V : O= 19.16% : 20.56% : 60.28%, statistically (see Table S5 for detailed EDX statistics). We note that EDX quantification of light-weight oxygen might be less accurate. More accurate XRF analysis confirmed Zn : V = 49.32% : 50.68% with a deviation of 0.86% (Figure S4B). Our combined local-probe EDX and macroscopic XRF analysis confirmed the elemental uniformity and composition of $ZnVO_3$.

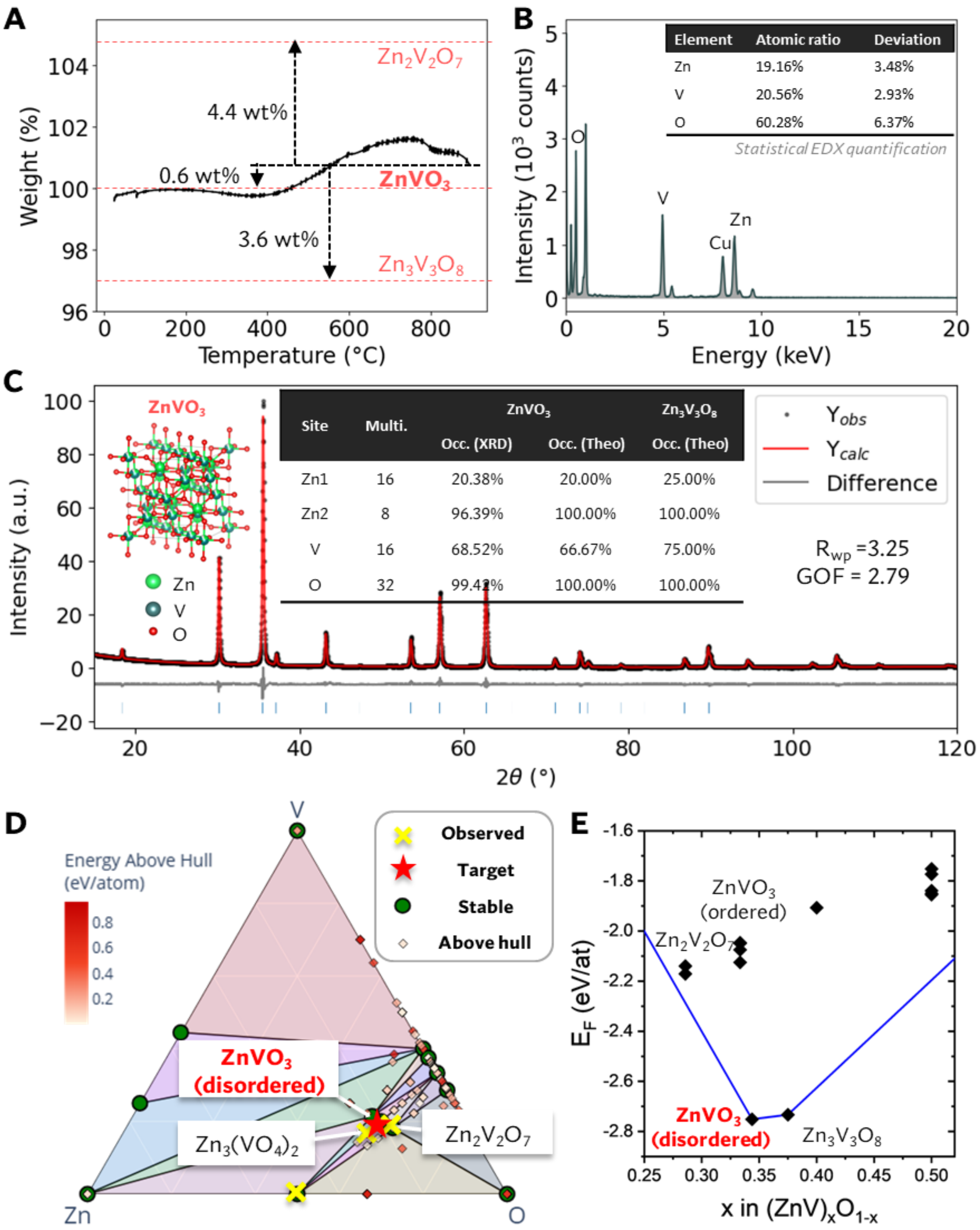

**Figure 4. XRD analysis of ultrafast synthesis products targeting $ZnVO_3$ via synthesis planned precursors.** (A) Thermogravimetric analysis (TGA) plot of $VO_2$ + ZnO reaction in inert atmosphere; (B) STEM-EDX spectrum and quantification statistics of the spinel $ZnVO_3$; (C) XRD refinement suggesting a partially disordered spinel $ZnVO_3$ structure. Site occupancy statistics averaged from five individual samples; (D) Ternary energy landscape of Zn-V-O overlaid with disordered $ZnVO_3$, (E) Energy diagram through Zn:V = 1:1 binary line.

Therefore, it is reasonable to consider the existence of spinel $ZnVO_3$ with possible interstitial oxygen atoms or metal vacancies. Further ESR analysis focusing on unpaired oxygen electrons revealed a flat spectrum (Figure S4A). Since unpaired oxygen electrons would resonate with a magnetic field, the absence of such resonance indicates that interstitial oxygen or oxygen vacancy is unlikely. XRD refinement was performed in Figure 4C to gain further insight into the unit cell structure and the potential presence of metal vacancies. $ZnVO_3$ exhibits a lattice parameter of 8.3959 Å. The lattice parameters of $ZnV_2O_4$, $Zn_3V_3O_8$, and $Zn_2VO_4$ are 8.4086 Å, 8.4005 Å, 8.3950 Å, respectively, displaying a decreasing trend as site disorder increases. The relatively small unit cell of $ZnVO_3$ suggests the potential presence of site disorder. It is observed that the octahedral site co-occupied by Zn1 and V is not fully occupied by metal atoms (25% Zn1 and 75% V when fully occupied), while the tetrahedral Zn2 site is nearly fully occupied. The refined occupancies from XRD (20.38% Zn1 and 68.52% V) closely resemble those of $ZnVO_3$ (20% Zn1 and 67% V) rather than $Zn_3V_3O_8$ with a similar spinel structure (25% Zn1 and 75% V), which is consistent with the earlier TGA studies. Five individual samples were prepared to reduce randomness in site disorder refinement (Table S6, detailed statistics), confirming a metal spinel $Zn_{2.74\pm0.058}V_{2.74\pm0.058}O_{7.95\pm0.049}$ structure. The refined stoichiometries are within reasonable measurement threshold, suggesting the successful synthesis of $ZnVO_3$ in a metal deficient spinel phase. Further negative refinement tests against existing Zn-V-O spinel phases were conducted (Figure S10). The refinement for $ZnVO_3$ was marginally better than that for $ZnV_2O_4$ and noticeably better than those for $Zn_3V_3O_8$ and $ZnV_2O_4$. Considering the differences in unit cell size and refinement quality, it is more reasonable to compare $ZnVO_3$ with the other spinel phases tested. The octahedral site deficiency observed in $ZnVO_3$ is analogous to that in $FeCrO_3$ (ICSD-196149). Both $ZnVO_3$ and $FeCrO_3$ exhibit a reduction in unit cell size relative to their parent compounds, $ZnV_2O_4$ and $FeCr_2O_4$, with shrinkages of 0.08% and 1.07%, respectively. The site disorder in these compounds contrasts slightly with that observed in other $ABO_3$ spinels. For example, $CdSnO_3$ (ICSD-23419) features deficiencies in both the octahedral and tetrahedral sites, resulting in a 0.88% unit cell expansion compared to $CdSn_2O_4$. $AlVO_3$ presents only a tetrahedral site deficiency, leading to an apparent 2.8% lattice expansion relative to $AlV_2O_4$. A more detailed comparison is attached in Table S7.

Given the discovery of the disordered spinel $ZnVO_3$, it becomes essential to assess the thermodynamic stability of spinel structures that remains stable during experimental exploration. However, disordered structures were not included in the original convex hull construction, due to computational challenges. We further constructed supercells of partially disordered spinel $ZnVO_3$ and $Zn_3V_3O_8$ structures for further DFT calculations (see Methods for details). Figure 4D presents the Zn-V-O ternary phase diagram, revealing new convex hull surfaces. $ZnVO_3$, previously considered metastable in an ordered *$P2_1/c$* structure during preliminary DFT screening (Figure S2), now resides on the hull surface. An energy diagram in Figure 4E focused on the Zn:V = 1:1 ratio further delineates that the hull surface is redefined by the lowest energy structure $ZnVO_3$ (disordered). Therefore, it is reasonable to assume higher stability from $ZnVO_3$ than $Zn_3V_3O_8$ considering their close stoichiometry and formation energy difference. The results are in excellent agreement with prior experimental studies, confirming the successful synthesis of thermodynamically stable $ZnVO_3$, crystallized in a metal-deficient spinel structure, as predicted by our in-silico approach.

***Case Study 2: High-throughput experimental synthesis targeting the predicted composition $YMoO_3$***

We next move to the synthesis exploration by focusing on empirically planned 'Y + $MoO_3$' precursors. In Figure S5A, the XRD pattern shows the temperature series of ultrafast synthesized products from Y and $MoO_3$ precursors in Ar environment. Careful deconvolution in Figure S5B reveals a coexistence of Mo, $MoO_2$, and $Y_2Mo_2O_7$ for synthesis up to 900 °C. The composition is consistent with that of the ball-milled precursors, suggesting that the reaction may require higher temperatures to initiate. At the next experimentally accessible temperature 1100 °C, multiple Y-Mo-O phases begin to appear. Notably, $Y_{2.5}MoO_6$ takes prominence over $Y_2Mo_2O_7$ at 1100 °C, followed by $Y_4Mo_4O_{11}$ at 1300 °C, $Y_4Mo_4O_{11}$ alongside $Y_{2.5}MoO_6$ at 1500 °C, and $Y_{2.5}MoO_6$ alongside $Y_2Mo_2O_7$ at 1700 °C. All major diffraction peaks were indexed and preliminarily refined without any sign of additional new phases. Further visualization of phases observed during ultrafast synthesis over the ternary Y-Mo-O phase diagram (see yellow crosses in Figure S5C) suggest that the reaction tends to be dominated by the formation of competing ternary YMO phases. To ascertain whether the YMO phases originate from the decomposition of high-temperature phases, *in situ* XRD synthesis from the same precursors was also conducted. A consistent XRD pattern (Figure S6A) persisted during cooling from 1000 °C, suggesting no formation of high temperature metastable phase. However, due to the limitations of our apparatus, *in situ* observations higher than 1000 °C remain unattainable.

Given that the combination reaction approach yields multiple competing Y-Mo-O phases, we next explored the more selective metathesis pathways predicted via the reaction network as discussed earlier. We focused on precursor sets from two lowest cost Cl-based metathesis pathways (a) 3 $Na_2CO_3$ + 2 $YCl_3$ + 2 $MoO_2$, and (b) $Na_2MoO_4$ + 2 YClO + Mo. Temperature series of XRD patterns in Figure 5A and phase evolution plot in Figure 5B depict the temperature series for the '$Na_2CO_3$ + $YCl_3$ + $MoO_2$' route. Detailed deconvolution revealed precursor reactions initiating within 300 - 500 °C forming binary oxides. Ternary products $Y_{2.5}MoO_6$ with some $Y_4Mo_4O_{11}$ increased as Mo- and Y-based precursors were depleted as temperature increases. New unassigned diffraction intensities around 2θ values of 21 and 24 degrees (checked against the PDF database; see black dashed boxes in Figure 5A) emerged from 1100 °C and peaked at 1200 °C (relevant composition represented by the highest new diffraction, shown as purple triangled line in Figure 5B), indicating the possibility of a new phase. Further XRD collected for 8h using extended scan range of the 1200 °C sample unveiled additional new diffractions below 20 degrees (see black dashed boxes from the XRD refinement in Figure S5J). On the other hand, the precursor set '$Na_2MoO_4$ + 2 YClO + Mo' (Figure S5G-I), predominantly produced $YMoO_4$ from 500 – 800 °C, succeeded by $Y_{2.5}MoO_6$ at temperatures exceeding 900 °C. Weak unassigned new diffraction intensities at 2θ values of around 34 and 40 degrees (see black dashed boxes in Figure S5G XRD patterns and Figure S5K XRD refinement of the 1200 °C synthesized sample) emerged starting at 1000 °C and peaking at 1200 °C. In both pathways, the occurrence of competing phases (see yellow crosses in Figure 5C and S5I) was largely reduced, giving concurrent formation of $Y_{2.5}MoO_6$ and new diffraction peaks that might be attributed to a new phase.

Pathways via Na-F and Na-Br were also analyzed and compared to understand the differences between these reactions. The XRD patterns and phase evolution of the Na-F based pathway are detailed in Figure S6B and S6C. As anticipated, YOF and NaF formed at 700 °C and 300 °C, respectively, potentially facilitating the progression toward the target. However, NaF was consumed by the competing $Na_3(MoO_4)F$, as identified during the phase evolution (Figure S6C), limiting the target reaction. The formation of $Na_3(MoO_4)F$ may be related to a direct reaction between the $Na_2MoO_4$ in precursor and the formed NaF, which inhibits the formation of $NaMoO_2$ via $Na_2MoO_4$ + Mo. The Na-Br-based pathway (Figure S6D-E) exhibited a distinct profile. In the reaction, carbonates, $Y_2MoO_6$, and molybdenum suboxides are the primary products from 100 °C, with 15 known phases observed as the temperature increased, suggesting a far more complicated reaction set beyond predictions. Given the high degree of mixing, interpreting the XRD pattern and discerning whether such new diffraction intensities stem from an unindexed phase or alternate sources becomes particularly challenging. The discovery of new diffraction peaks during Na-Cl based metathesis reaction indicates the synthesis of one or more potentially undiscovered phase(s), warranting further chemical and structural investigation.

We then used localized S/TEM based techniques to analyze on the products from the '$Na_2CO_3$ + $YCl_3$ + $MoO_2$' metathesis reaction. Figure 5D-F shows results from the crystals that exhibit a composition close to the target $YMoO_3$ composition within the as-synthesized sample. They crystallize in a rod-type morphology with the competing $Y_{2.5}MoO_6$ phase nearby (Figure 5D, lower right). Observed from EDX mapping (Figure 5E), elemental ratios were acquired and used to differentiate the competing $Y_{2.5}MoO_6$ phase against the phase with a composition close to $YMoO_3$, separately circled in blue dashed lines and white solid lines. To obtain a more accurate quantification, multiple regions were analyzed, which confirmed a statistical elemental ratio Y : Mo : O = 20.72% : 18.82% : 60.46% (Figure 5F). Elemental information was further confirmed using a different TEM, which revealed consistently similar quantification (Table S5). HR-TEM image with an inset DF-TEM image and a corresponding selected area diffraction (SAD) pattern is displayed in Figure 5D. The highly periodic patterns suggest that the new phase comprises sub-micron rod-like single crystals, consistent with the dark field TEM (DF-TEM) image obtained from selected low-index diffraction spots. Therefore, this crystal is closely related to the target composition $YMoO_3$.

Microcrystal electron diffraction (microED) was subsequently carried out to perform localized *ab initio* structure solution within the powder mixture for the $YMoO_3$-like crystals, following the workflow illustrated in Figure S7. Briefly, crystals exhibiting a composition close to 1:1:3 via STEM-EDX were identified on the TEM holder. A narrow electron beam (approximately 40-200 nm) was then used to optimize the diffraction signal locally on these crystals, followed by continuous rotation for electron diffraction data collection. The angle-resolved diffraction series enabled the reconstruction of the 3D reciprocal space, when local symmetry and the unit cell were determined via systematic absence criteria and reciprocal lattice dimensions. The final structural solution was conducted in Olex2 software, similar to the approach used in single crystal X-ray diffraction. Figure 5G shows the reconstructed reciprocal space from microED data. $Y_4Mo_4O_{11}$, with an orthorhombic *Pba2* unit cell of a = 10.7150 Å, b = 15.9032 Å, c = 5.6645 Å (Figure 5H), was proposed as the best solution. Structural comparison with existing *Pbam* $Y_4Mo_4O_{11}$ polymorph[49] revealed minor atomic position deviations and a lack of centrosymmetry in the unit cell (Figure 5H, along the b-axis). Given that microED typically lacks the intensity contrast needed to resolve partial disorder (e.g., vacancies, site disorder), $YMoO_3$-like crystals adhere to the symmetry and atomic positions of the resolved $Y_4Mo_4O_{11}$ structure, but not the composition.

To validate the structure on a macroscopic level, a HR-XRD pattern (Figure 5I) collected from samples washed using DI water for the removal of NaCl was refined according to the microED resolved structure and other known phases, indicating a high degree of matching. Further characterization to confirm its composition and potential site disorder is desirable, but the difficulties in purification have limited these efforts. The new peaks observed during screening between 10-25 degrees (Figure S5J) in samples prior to washing are potentially related to water-soluble Na- or Cl-based compounds not cataloged in the PDF/ICSD database. On the other hand, the products from '$Na_2MoO_4$ + Mo + YClO' reaction was examined via S/TEM in Figure S8. A flower-like Y-Mo-O nanocrystal was identified whose structural information requires advanced nanoscale structure solution method to elucidate.

Further ad-hoc DFT investigations were carried out including partially disordered $Y_{2.5}MoO_6$ absent in earlier DFT screening (Figure S9, Y-Mo-O ternary energy landscape), showing a new convex hull surface redefined by thermodynamically stable $Y_{2.5}MoO_6$ and $Y_4Mo_4O_{11}$ phases. The consistent appearance of $Y_{2.5}MoO_6$ and $Y_4Mo_4O_{11}$ is due to incomplete synthesis planning, which is limited by the lack of thermodynamic data for disordered structures in openly accessible thermodynamic databases. In conclusion, the synthesis efforts targeting $YMoO_3$ resulted in two compositionally similar crystals: a rod-like crystal with a structure resembling $Y_4Mo_4O_{11}$ from the '$Na_2CO_3$ + $YCl_3$ + $MoO_2$' reaction and a flower-like nanocrystal from the '$Na_2MoO_4$ + Mo + YClO' reaction.

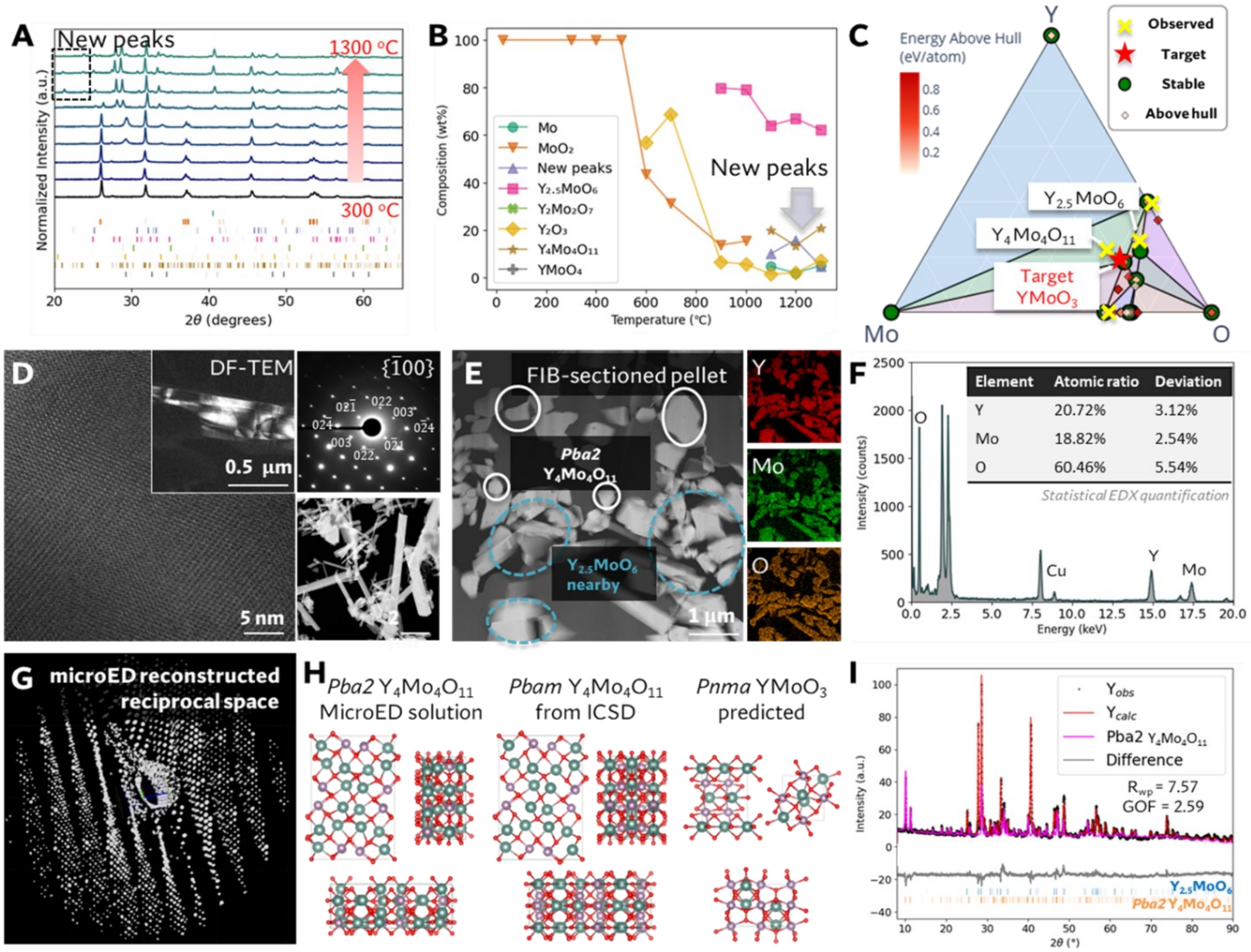

**Figure 5. XRD analysis of ultrafast synthesis products targeting $YMoO_3$ via synthesis planned '$Na_2CO_3$ + $YCl_3$ + $MoO_2$' reaction with detailed characterizations of rod-like $Y_4Mo_4O_{11}$ with elemental ratio close to 1:1:3.** (A) Temperature series of XRD patterns, (B) Phase evolution resolved via XRD screening, and (C) Ternary phase diagram overlayed with competing phases observed, acquired from the Materials Project; (D) HR-TEM with inset DF-TEM image showing a rod-type $Y_4Mo_4O_{11}$ single crystal, SAD image from $\{\bar{1}00\}$, and HADDF-STEM overview image; (E) Cross-sectioned HADDF-STEM image of a powder pellet with corresponding EDX maps. (F) EDX spectrum over target crystals and quantification statistics; (G) microED reconstructed reciprocal space and (H) *Pba2* structure solution compared with relevant structures (Top left: c-axis, top right: b-axis, bottom: a-axis.); (I) high resolution XRD spectrum collected on a NaCl-free sample fitted with microED solved structure and known phases. Characterizations of the flower-type nanocrystals produced via '$Na_2MoO_4$ + Mo + YClO' metathesis.

## Discussion

Through meticulous candidate selection, synthesis planning, rapid synthesis mechanism, and in-depth structural and compositional characterizations, we have demonstrated the effectiveness of our end-to-end paradigm in approaching target composition, while circumventing many potential competitors. We found that $ZnVO_3$ can be successfully synthesized following our predicted pathway, which crystallizes into a metal-deficient spinel structure, a point never reported on Zn-V-O phase space. We consider this a step forward, considering most random compositions do not exist on the phase diagram (except for alloys). Further iterations of more detailed DFT investigation within the Zn-V-O ternary phase diagram unveiled a new convex hull, suggesting that this metal-deficient spinel $ZnVO_3$ structure is favorable compared to hypothetical ordered polymorphs of $ZnVO_3$. The exploration targeting $YMoO_3$ revealed two different crystals close to the target $YMoO_3$ composition, including a rod-like crystal with a structure close to $Y_4Mo_4O_{11}$.

A major limitation is the treatment of partial disorder, which affects over 50% of ICSD entries but is often neglected in generative models, limiting their predictive reach. As shown in a case study (Table S8), the presence of partially disordered neighboring phases statistically lowers the likelihood of hypothetical structures being realized. Sparse thermodynamic data for disordered systems further hampers accurate prediction. Progress in high-throughput thermodynamic modeling and improved databases for disordered materials would enhance reliability.

Another challenge lies in retrosynthesis: current reaction network-based methods show potential for predicting compositions but fall short when reaction kinetics are considered. Structure-targeted synthesis thus remains difficult. In parallel, existing diffraction and elemental analysis techniques have limited accuracy especially for disordered materials and non-homogeneous mixtures, underscoring the need for more advanced characterizations. Looking forward, we anticipate that developments in both computational and experimental methods will help close the experiment-theory gap, broadening the reach of inorganic materials discovery and accelerating the realization of materials by design.

## Methods

### *Machine learning synthesizability and date-driven oxidation state probability filter*

The Synthesizability Score (SC) model we developed previously[9] was employed to evaluate the synthesis potential of 203 non-ICSD $ABO_3$ candidates downloaded from the Materials Project[39]. The model structure and hyperparameters in the original SC model are employed to ensure a similar level of performance. The specific SC model used in this work is trained using a 5-fold cross validation (CV) method, where the SC model in each fold is trained and evaluated using the dataset separated into 80% for training and 20% for testing. SC scores are the average of prediction from all five SC models (trained in 5-fold CV). The oxidation state probability is estimated by an OSP model[16] by comparing the statistical likelihood of a compound taking a particular combination of elemental oxidation states.

### *DFT calculations*

The crystal structure prediction of $ZnVO_3$ has been performed using the Crystal structure AnaLYsis by Particle Swarm Optimization (CALYPSO) package[50], within which the total energy calculations are implemented by the Vienna Ab initio Simulation Package (VASP)[51]. The population size of 80 structures was employed in each generation of the particle swarm optimization (PSO) evolution. Each population was comprised of 60% PSO-generated structures and 40% randomly generated ones. Projector augmented wave (PAW) pseudopotentials are used to describe the ion-electron interaction in DFT calculations. An energy cutoff of 520 eV for the plane waves was used[52]. The Perdew-Burke-Ernzerhof (PBE) functional within the generalized gradient approximation (GGA) is employed to approximate the exchange-correlation of valence electrons[53]. The integral over the Brillouin zone was done by sampling using a Γ-centered $k$-mesh with a spacing of 0.15 $Å^{-1}$. The Monkhorst-Pack scheme is used to sample $k$-points in the Brillouin zone. The tetrahedron method with Blöchl corrections is employed for determine the orbital occupancy. The convergence criteria of $10^{-7}$ eV for total energy and $10^{-4}$ eV $Å^{-1}$ per atom for force were used.

For $Y_{2.5}MoO_6$, a partially disordered structure (*Immm*) experimentally indexed from PDF database is included. Considering that the site occupancy is inter-correlated, we constructed 2 distinct virtual cells and relaxed them independently. Further DFT calculation using r$^2$SCAN method on Y-Mo-O system was carried out using settings from the Materials Project[54]. The compounds for convex hull construction were queried from the Materials Project, with the search limited to those calculated via the same method only. For Zn-V-O mixed spinel structures, a partially disordered structure ($Fd\bar{3}m$) experimentally indexed is included. *sqsgenerator,* a cluster expansion-based programme where randomness is measured by Warren-Cowley short range parameters[55] was used to treat disorder and generate the supercell. The spinel $Zn_3V_3O_8$ unit cell includes sites that are half-occupied by Zn/V ions, and a quasirandom supercell of 2×1×1 unit cells of 112 atoms (24 V, 24 Zn, 64 O) is generated using the *sqsgenerator*[56] to represent $Zn_3V_3O_8$. The spinel $ZnVO_3$ unit cell has additional metal vacancies, thus a supercell of 108 atoms (22 V, 22 Zn, 64 O) is generated similarly.

### *Data-driven synthesis planning*

Synthesis recipes were calculated following the methodology and approach outlined in this reference[42]. The chemical system considered the 3 target elements (Y, Mo, O) and 14 additional elements (Li, Na, K, Mg, Ca, Sr, F, Cl, Br, S, N, C, B, and H), commonly found in precursors for synthesis. The energies are modeled as Gibbs free energies at a temperature of T=650 °C using the ML-based Gibbs energy descriptor from Bartel *et al.*[57]. The reaction cost is determined as the weighted average of the reaction energy ($\Delta G_{rxn}$), primary competition score ($C_1$), and secondary competition score ($C_2$), with weights of 0.1, 0.45, and 0.45, respectively. Reactions with lower (more negative) costs are considered more favorable. The synthesis recipes of heuristic combination reactions are calculated in a similar manner, without additional elements nor considering competing reactions. Reactions with lower reaction energy are considered more favorable.

### *Ultrafast synthesis*

Ultrafast synthesis was carried out according to the workflow illustrated in Figure 3A. The precursors are weighed according to their stoichiometric ratio with a 3g total weight and 1mg accuracy. The precursors are then sealed into an Ar-filled stainless steel jar and milled at 300 rpm for 3 hours in a planetary ball mill system. 70 mg mixed powders are then transferred into a 1cm wide graphite paper envelope which acts as a reaction chamber during heating later. Note that graphite paper under high temperature tends to be reducing. Pt envelope may be used to circumvent this. The envelope is then sandwiched into a pair of carbon filament connected to a power source Keithley 2260B (DC 80V, 40A) for solid state synthesis at a designated temperature for 30 seconds in an Ar-filled glove box. The temperature is calibrated according to the 15s temperature measured by an IR probe. The system is capable of heating from 100 °C to 2000 °C coupled with this Keithley power supply. Owing to the highly thermally conductive nature of the carbon-based reaction chamber, heat transfer and dissipation are accelerated contributing to rapid ramping and cooling during synthesis at $10^3$ K/s rate (Figure 3C and 3D). The heating and cooling rates are determined by the largest temperature derivative. Experimentally, if the difference between environmental temperature and target temperature is significant (e.g. 1000 K difference), then the heating and cooling rate are larger; conversely, when the difference is small (e.g. 200 K), the rate is lower due to limited heat dissipation capacity.

A comparative study of conventional furnace synthesis and ultrafast synthesis for $ZnVO_3$ in an inert atmosphere (Figure S3A and S3B, orange XRD curves) revealed that both methods produced similar products: the ' $VO_2$+ZnO' reaction resulted in pure spinel phase formation, while ' $V_2O_5$+ Zn' yielded a dominant spinel phase with minor ZnO precipitates. These findings demonstrate that, in this case, ultrafast synthesis achieves the same phase as conventional synthesis. We note that ultrafast synthesis does not function in open air due to potential carbon filament oxidation.

*In situ* XRD synthesis was carried out on a ceramic stage with a graphite cover (Anton Paar, DHS1100) to maintain vacuum status. The powder is compressed into a 13mm wide and 1mm thin pellet to ensure proper X-ray coverage and heat conduction. Conventional synthesis was carried out using a tube furnace. The powders were loaded into the furnace and annealed for 30 mins before cooling down in their respective atmosphere. After heating, the powder was grounded using a mortar and pestle for 5 mins before further characterization.

***Characterization and analysis***

SEM images were taken on a JEOL JSM-7800F Prime with Oxford Ultim Max 170mm$^2$ EDX system. Focused ion beam cross-sectioning was carried out via a ZEISS Crossbeam 540 with Ga-ion source. TEM was conducted on JEOL ARM-300F with JEOL JED-2300T 100mm$^2$ (~0.98 sr) windowless SDD EDX detector and JEOL 2100F with Oxford Ulti Max 100mm^2 (~0.7 sr) windowless detector. FFT pattern was processed in Gatan Digital Micrograph. *ex situ* XRD was carried out on a Bruker D8 Advanced diffractometer under coupled theta-2theta geometry with a Cu Kα X-ray source operated at 40kV and 40mA ($\lambda_{CuK\alpha}$ = 1.54056 Å) and a beam knife for low-background low-angle signal. Rietveld refinement was performed on workstations licensed with TOPAS V6. *In situ* XRD was carried out on a Bruker D8 Discover diffractometer with an Anton Paar DHS 1100 heating stage, configured as under coupled theta-2theta geometry with a Cu Kα X-ray source operated at 40kV and 40mA ($\lambda_{CuK\alpha}$ = 1.54056 Å). microED data was collected using Rigaku XtaLAB Synergy-ED, indexed via CrysAlis$^{Pro}$, and solved via OlexSys Olex 2. Thermo-analysis was conducted using a TA Q500 Thermogravimetric Analyzer. ESR analysis were carried out using a JEOL JES-X320 ESR spectrometer.

**Resource availability**

**Lead contact:**

- Requests for further information and resources should be directed to and will be fulfilled by the lead contact, Dai Haiwen (haiwen.dai@ntu.edu.sg)

**Materials availability:**

- This study did not generate new unique reagents. Reagents in the study are available for direct order from Sigma Aldrich.

**Data and code availability:**

- The code for DTMA is available in the Zenodo (https://doi.org/10.5281/zenodo.15715403). Further updates can be found on GitHub (https://github.com/Kedar-Materials-by-Design-Lab/Design-Test-Make-Analyze/).
- All data is available in the main text and supplemental information. Original data accessible in the Zenodo (https://doi.org/10.5281/zenodo.15715394).
- Any additional information required to reanalyze the data reported in this paper is available from the lead contact upon request.

## Acknowledgments

**Funding statement:** The authors gratefully acknowledge financial support from

- Agency for Science, Technology and Research Grant No. M24N4b0034 (KH).
- Agency for Science, Technology and Research Grant No. 192D8230 (FW).
- Agency for Science, Technology and Research Grant No. C210112054 (ADH).
- National Research Foundation Singapore Grant No. NRF-NRFF13-2021-0011 (KH).
- National Research Foundation Singapore Grant No. NRF-CRP25-2020-0002 (KH).
- the Materials Project program (KC23MP) under Contract No. DE-AC02-05CH11231 (KAP).

## Author contributions

Conceptualization: KH, KAP. Methodology: HD, JRG. Software: MJM, APC, WN, RZ, MT. Investigation: HD, SM, CS, SDP, CZ, WGS, BNT, PM, FW, ADH, SH, HS. Supervision: KH,

KAP, ML, CWL. Visualization: HD, MJM, APC, WN, RZ, CS. Writing - original draft: HD. Writing - review & editing: All authors reviewed and edited the manuscript.

**Declaration of interests:**

The authors declare that they have no conflict of interests.

**Supplemental information**

Supplemental Information is available in a separate file including Note S1, Figure S1 to S10 and Table S1 to S8.

# Supplemental Information for

# Data-driven Design–Test–Make–Analyze Paradigm for Inorganic Crystals: Ultrafast Synthesis of Ternary Oxides

Haiwen Dai[1], Matthew J. McDermott[2], Andy Paul Chen[1], Jose Recatala-Gomez[1], Wei Nong[1,3], Ruiming Zhu[1,4], Maung Thway[1], Samuel Morris[1], Christian Schürmann[5], Shreyas Dinesh Pethe[1], Chenguang Zhang[1], Wuan Geok Saw[6], Bich Ngoc Tran[6], Pritish Mishra[1,7,8], Fengxia Wei[4], Albertus Denny Handoko[4], Sabrine Hachmioune[4,9], Haipei Shao[10], Ming Lin[4], Chong Wai Liew[6], Kristin A. Persson[11], and Kedar Hippalgaonkar[1,4,*]

This Supporting Information include Note S1, Figure S1 to S10, and Table S1 to S8.

**Note S1. A step-by-step guide for using the synthesizability (SC) and oxidation state probability (OSP) models for evaluation**

Inputs:

- SC model: Requires a DFT-relaxed CIF file (full crystallographic structure with unit cell and atomic coordinates).
- OSP model: Requires the chemical composition of the compound.

Workflow:

1. Start with a CIF file of the candidate compound. If the structure is not already relaxed, perform DFT relaxation to obtain a meaningful input.
2. Confirm the structure is not already in the Materials Project database (optional). One can use the StructureMatcher() function from the pymatgen package to check whether the structure exists in MP (v2022.10.28). If it does, we recommend using a leave-one-out approach retraining the model, since the pretrained model may have been trained on this structure.
3. Predict synthesizability (SC). Input the relaxed CIF file into the pretrained synthesizability model. The output is a synthesizability score ranging from 0 to 100%. Model access and usage instructions are available on GitHub: https://github.com/Kedar-Materials-by-Design-Lab/Design-Test-Make-Analyze/
4. Predict oxidation state probability (OSP). Input the chemical composition into the OSP model script. The model returns a raw probability, which should be normalized by the highest probability compound within the same chemical system. For example, if $AB_2O_4$ has a raw probability of 0.5 and $ABO_3$ has 0.2, the normalized OSP for $ABO_3$ is 0.2 / 0.5 = 0.4. This normalization enables comparison across different systems. The model and instructions are available at the same GitHub link above.
5. Rank candidates. Multiply SC × OSP to obtain a final score between 0 and 100%, which can be used to prioritize candidate materials similar to Table 1 of the manuscript.

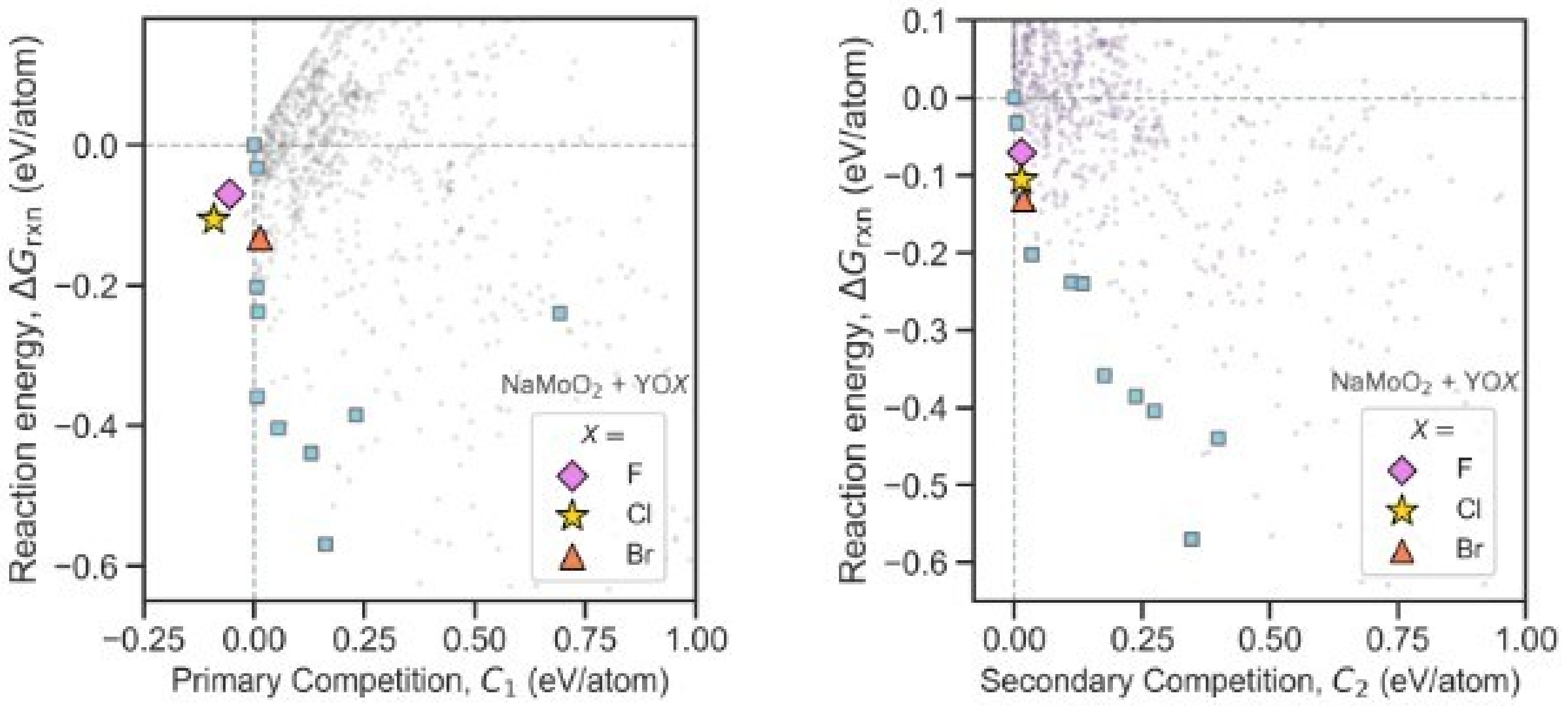

| Rank | Reaction | $\Delta G_{rxn}$ | C1 | C2 | Cost |
|---|---|---|---|---|---|
| 1 | $NaMoO_2$ + YClO -> $YMoO_3$ + NaCl | -0.1057 | -0.0917 | 0.014 | -0.0455 |
| 2 | $NaMoO_2$ + YOF -> $YMoO_3$ + NaF | -0.0701 | -0.0561 | 0.014 | -0.026 |
| 3 | 1.5 $Mo_9O_{26}$ + 4.667 $Y_2O_3$ -> $YMoO_3$ + 4.167 $Y_2(MoO_4)_3$ | -0.2023 | 0.0069 | 0.035 | -0.0014 |
| 4 | 0.5 $Y_2O_3$ + 0.3333 $K(MoO_2)_4$ -> $YMoO_3$ + 0.1667 $K_2Mo_2O_7$ | 0.0001 | 0.0001 | 0.000 | 0.0001 |
| 5 | $NaMoO_2$ + YBrO -> NaBr + $YMoO_3$ | -0.1319 | 0.0131 | 0.0181 | 0.0008 |

Unit of energy: eV/atom

**Figure S1. Data-driven synthesis planning via reaction network for $YMoO_3$ showing the thermodynamic driving forces and selectivities of 1278 synthesis reactions.**

A highly selective reaction will have a large thermodynamic driving force ($\Delta G_{rxn}$), small primary competition ($C_1$) and secondary competition ($C_2$). Reactions on the three-dimensional Pareto front are plotted as blue squares. Na-based ternary metathesis reactions are highlighted by anion (X=F, Cl, Br).

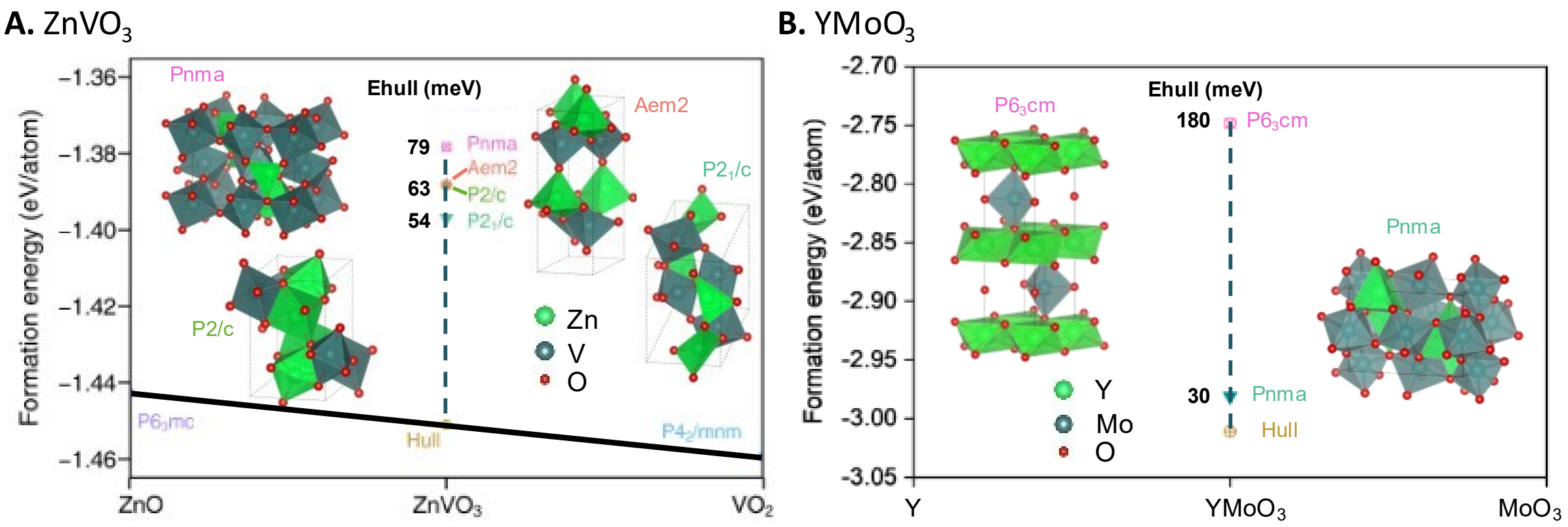

| Candidate | $E_{hull}$ (meV/atom) |
|---|---|
| $YMoO_3$ | 30 |
| $ZnVO_3$ | 54 |

**Figure S2. DFT-identified polymorphs of (A) $ZnVO_3$ and (B) $YMoO_3$.**

**Optimization of $ZnVO_3$:** The lowest-energy structure on the Materials Project has an $E_{hull}$ of 105 meV/atom indicating limited thermodynamic stability. Performing structural evolution via CALYPSO (Figure S1a) reveals an *Aem2* polymorph. Another structural evolution published recently identified additional two polymorphs *P2₁/c* and *P2/c* (*J. Phys. Chem. C* **127**, 11902–11910 (2023)). Relaxing all the low-energy polymorphs, we find that the most stable polymorph is *P2₁/c* space group with an $E_{hull}$ of 54 meV/atom, suggesting reasonable thermodynamic stability with respect to the competing decomposition products in the ternary space. Further details of $ZnVO_3$ structures are attached in Table S1 attached below.

**Optimization of $YMoO_3$:** The lowest $E_{hull}$ from the Materials Project stands at 30 meV/atom for a *Pnma* structure (Figure S1b). Although it is interesting to further optimize and explore lower energy polymorphs via theoretical means, such a small $E_{hull}$ indicates the potential accessibility of the target composition experimentally. Consequently, $YMoO_3$ and $ZnVO_3$, exhibit reasonably good stability and are promising candidates warranting further exploration through experimental means.

**Table S1.** Crystallographic comparison among various $ZnVO_3$ structures

| Entry | Unit cell | Atomic position | Structure description |
|---|---|---|---|
| mp-1016931<br>$ZnVO_3$<br>(DFT)<br>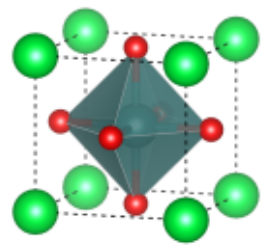 | P m -3 m (221)<br>a=3.7804 | Zn @ [1a] (0, 0, 0)<br>V @ [1b] (1/2, 1/2, 1/2)<br>O @ [3c] (0, 1/2, 1/2) | Polyhedra:<br>Corner-sharing [$VO_6$] octahedra and edge-sharing [$ZnO_{12}$] cuboctahedra<br><br>Type:<br>Cubic perovskite |
| mp-1378086<br>mvc-15883<br>$ZnVO_3$<br>(DFT)<br>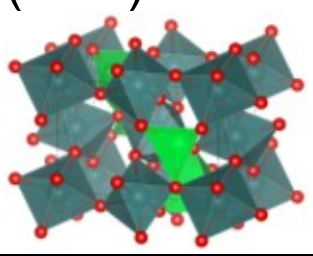 | P n m a (62)<br>a=5.1505<br>b=7.6435<br>c=5.1539 | Zn @ [4c] (0.0336, 1/4, 0.5069)<br>V @ [4a] (0, 0, 0)<br>O1 @ [8d] (0.1914, 0.5621, 0.6703)<br>O2 @ [4c] (0.4272, 1/4, 0.6138) | Polyhedra:<br>Tilted, distorted corner-sharing [$VO_6$] octahedra<br><br>Type:<br>Orthorhombic perovskite |
| P2/c<br>$ZnVO_3$ | P 2/c (13)<br>a=5.4942<br>b=5.0829 | Zn @ [4g] (0.2858, 0.4922, 0.3606)<br>V @ [4g] (0.2064, 0.9946, 0.1480) | Polyhedra:<br>Edge-sharing [$ZnO_6$] and [$VO_6$] octahedra; [$ZnO_6$] |
| (DFT)<br>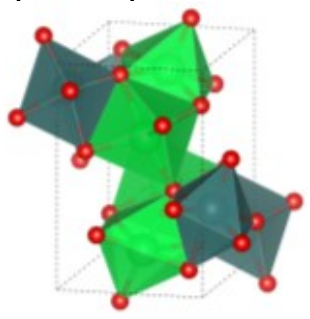 | c=7.6155<br>β=95.6182 | O1 @ [4g] (0.1493, 0.1443, 0.9107)<br>O2 @ [4g] (0.6645, 0.6849, 0.3984)<br>O3 @ [2f] (1/2, 0.1877, 1/4)<br>O4 @ [2e] (0, 0.2728, 1/4) | layer and [$VO_6$] layer stacking in alternating sequence along crystal b axis by sharing edges<br><br>Type:<br>Non-perovskite |
| $P2_1/c$<br>$ZnVO_3$<br>(DFT)<br>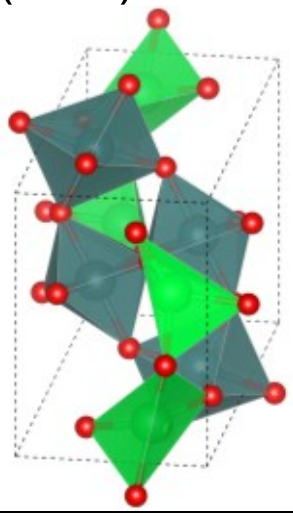 | P $2_1$/c (14)<br>a=6.6377<br>b=5.2772<br>c=7.0387<br>β=115.9986 | Zn @ [4e] (0.9634, 0.8020, 0.6336)<br>V @ [4e] (0.5620, 0.2545, 0.4378)<br>O1 @ [4e] (0.3016, 0.4096, 0.4005)<br>O2 @ [4e] (0.1161, 0.8266, 0.4398)<br>O3 @ [4e] (0.5612, 0.4288, 0.1876) | Polyhedra:<br>Corner/edge-sharing [$ZnO_4$] tetrahedra and [$VO_6$] octahedra; [$ZnO_4$] layer and [$VO_6$] layer stacking in alternating sequence along crystal *a* axis by sharing corners<br><br>Type:<br>Non-perovskite |
| Aem2<br>$ZnVO_3$<br>(DFT)<br>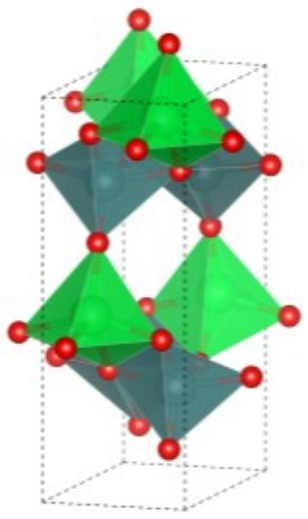 | A e m 2 (39)<br>a=6.6900<br>b=3.9586<br>c=10.8172 | Zn @ [4c] (0.8393, 1/4, 0.4478)<br>V @ [4c] (0.3305, 1/4, 0.2655)<br>O1 @ [4c] (0.6204, 1/4, 0.2956)<br>O2 @ [4c] (0.1504, 1/4, 0.4012)<br>O3 @ [4c] (0.2421, 1/4, 0.1198) | Polyhedra:<br>Corner/edge-sharing [$ZnO_5$] and [$VO_5$] pentahedra<br><br>Type:<br>Non-perovskite |

Refs: *J. Phys. Chem. C* **2023**, 127, 25, 11902–11910 (DOI: 10.1021/acs.jpcc.3c03583)

* Only free parameters and inequivalent Wyckoff positions are presented. Lattice length is in Å, angle in degree, and atomic coordinate in crystal fractional. Color codes for atoms: O in red, Zn in green, V in dark cyan, La in purple.

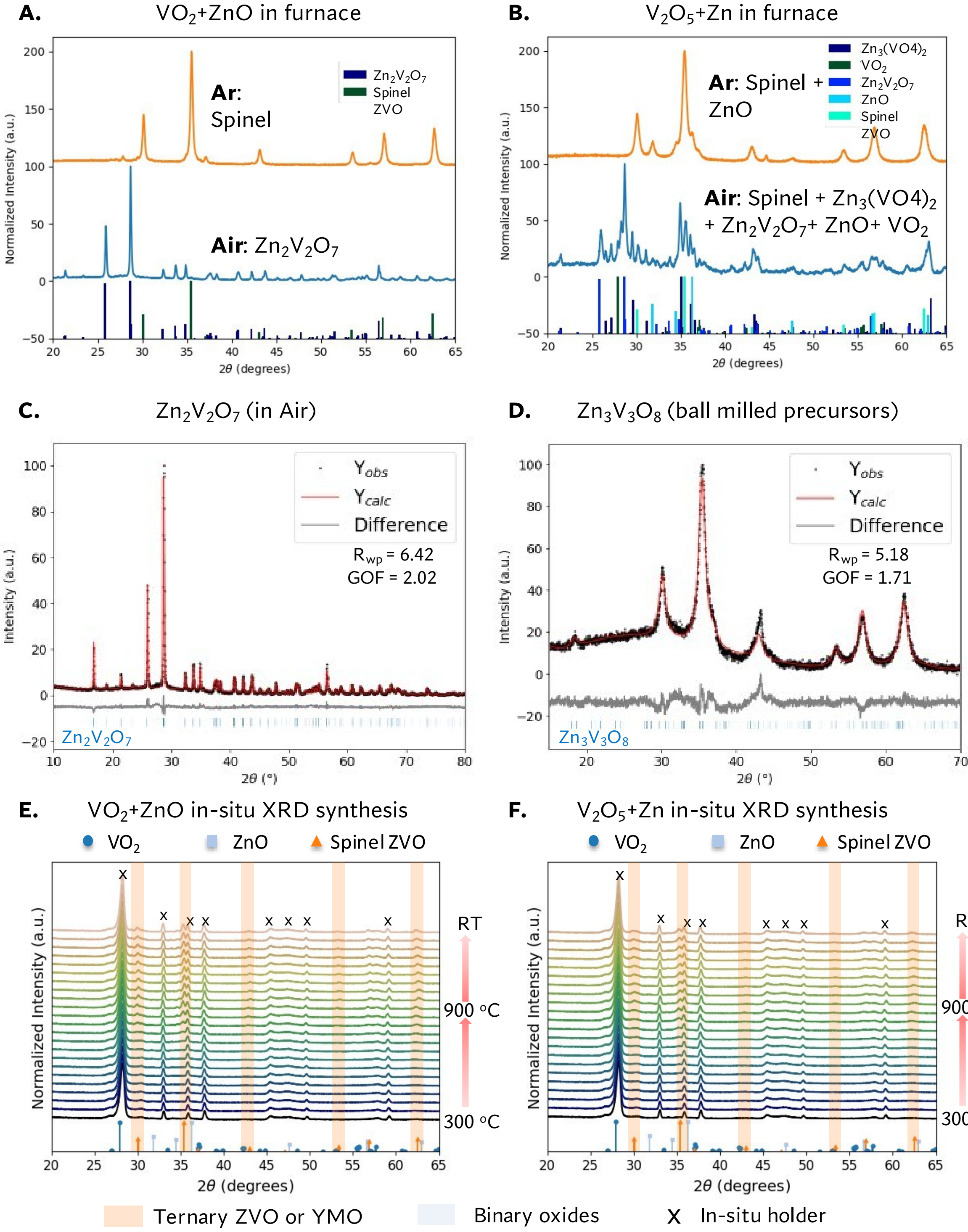

**Figure S3. XRD analysis towards $ZnVO_3$ synthesis.**

(A) $VO_2$ + ZnO and (B) $V_2O_5$ + Zn precursors in both Ar and air environment; XRD refinements of (C) sintered $Zn_2V_2O_7$ from $VO_2$ + ZnO in air, and (D) Ar ball milled $Zn_3V_3O_8$ from $V_2O_5$ + Zn. In-situ XRD synthesis of (E) $ZnVO_3$ via $VO_2$+ZnO, and (F) $ZnVO_3$ via $V_2O_5$+Zn. Spinel ZVO is always the major product and stable from high temperature to room temperature when targeting at $VZnO_3$. No high-temperature phase was found when targeting at $YMoO_3$.

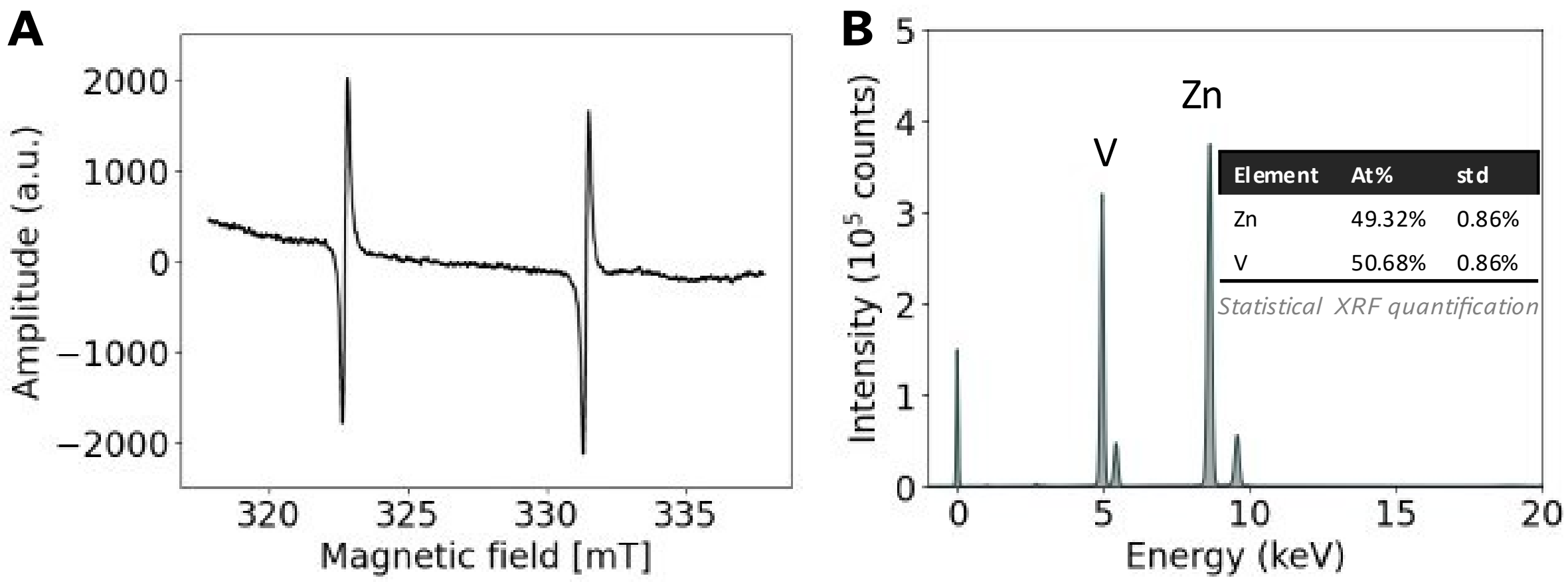

**Figure S4. Additional characterization of $ZnVO_3$**

(A) Electron spin resonance (ESR) spectrum with Mn markers. The 2 strong peaks from Mn are used for sample calibration. No other peak signals were observed. (B) XRF spectrum of pressed $ZnVO_3$ powder pellet with quantification statistics.

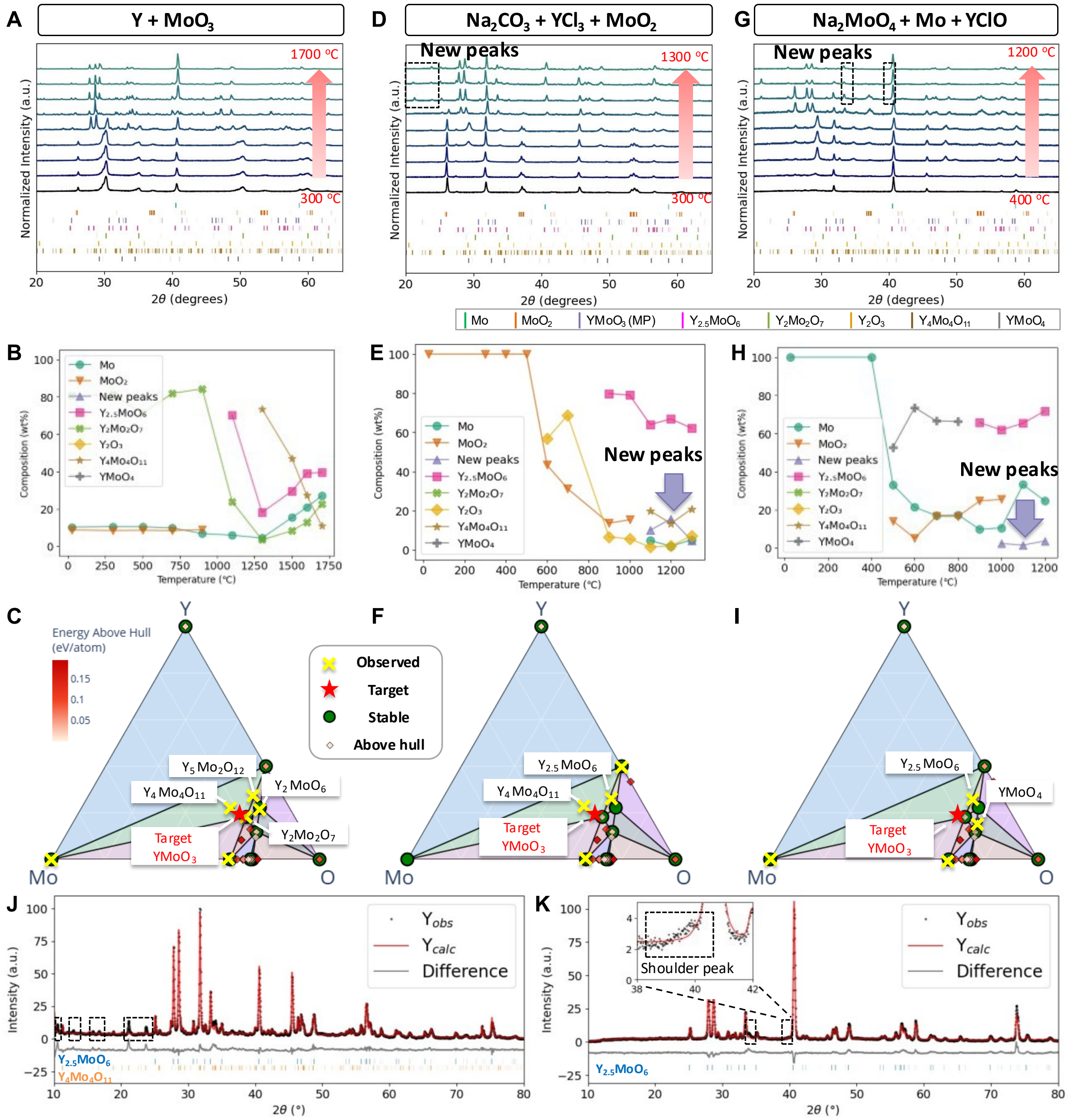

**Figure S5. XRD analysis of ultrafast synthesis products targeting $YMoO_3$ via synthesis planned precursors.**

Precursor set (A-C) 'Y + $MoO_3$', (D-F) '$Na_2CO_3$ + $YCl_3$ + $MoO_2$', and (G-I) '$Na_2MoO_4$ + Mo + YClO', with their temperature series XRD patterns (first row), phase evolution resolved via XRD refinement (second row), and ternary phase diagram overlayed with competing phases observed (third row). XRD pattern (collected for 8h using extended scan range) and refinement of the 1200 ℃ metathesis synthesized samples via (J) '$Na_2CO_3$ + $YCl_3$ + $MoO_2$' and (K) '$Na_2MoO_4$ + Mo + YClO' pathways. New diffraction intensities that cannot be attributed to known phases in PDF database (see black dashed boxes) suggests potential existence of a new phase. Compositions of potential new peaks in (E) and (H) are represented by their diffraction intensities. Both metathesis pathways helped reduce the formation of competing phases. The phase diagrams from the Materials project.

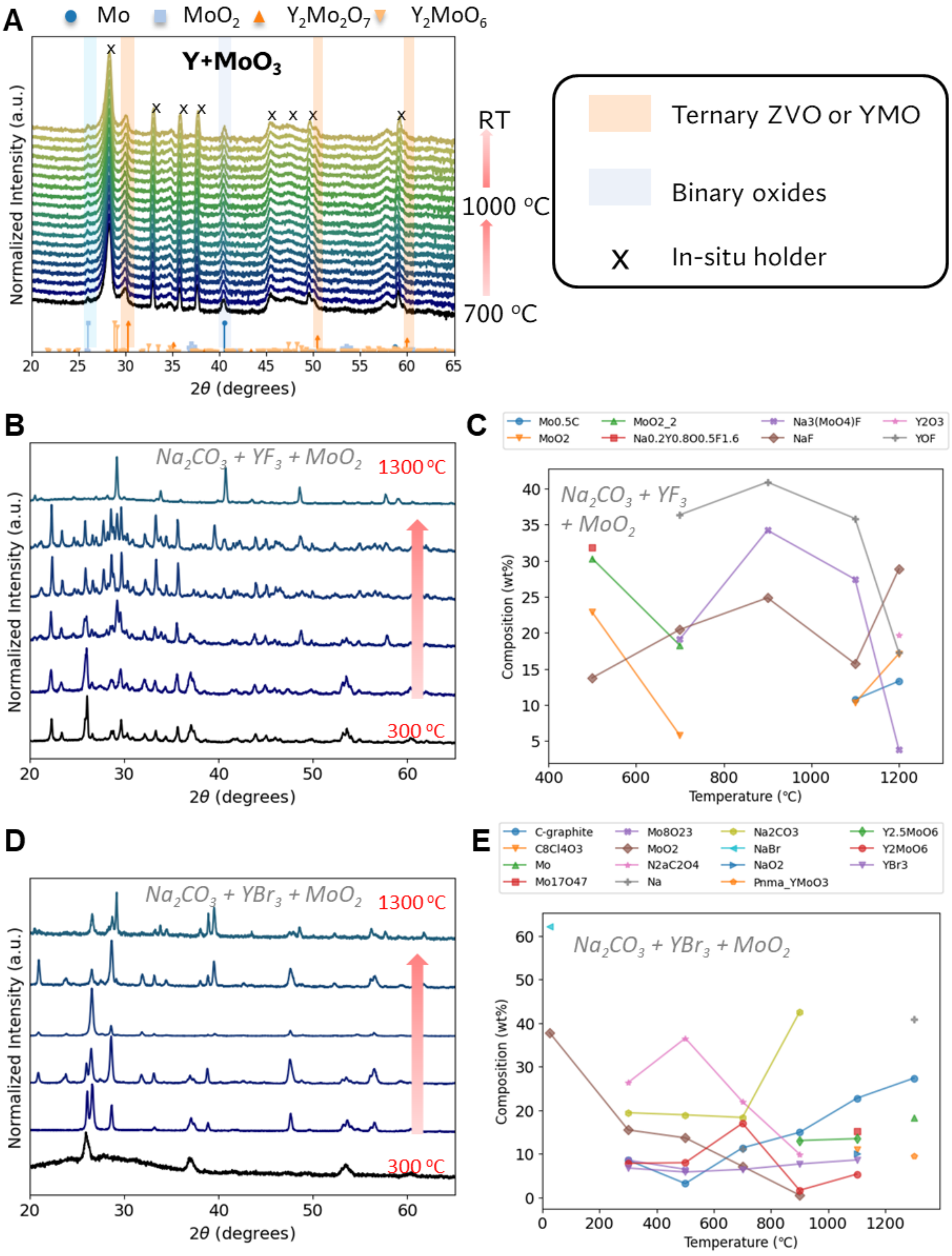

**Figure S6. XRD analysis towards ZnVO3 synthesis.**

(A) In-situ XRD synthesis targeting at $YMoO_3$; No high-temperature phase was found when targeting at $YMoO_3$; (B-C) $Na_2CO_3$ + $YF_3$ + $MoO_2$, and (D-E) $Na_2CO_3$ + $YBr_3$ + $MoO_2$. NaCl was removed from the phase evolution plot being a predicted byproduct of metathesis.

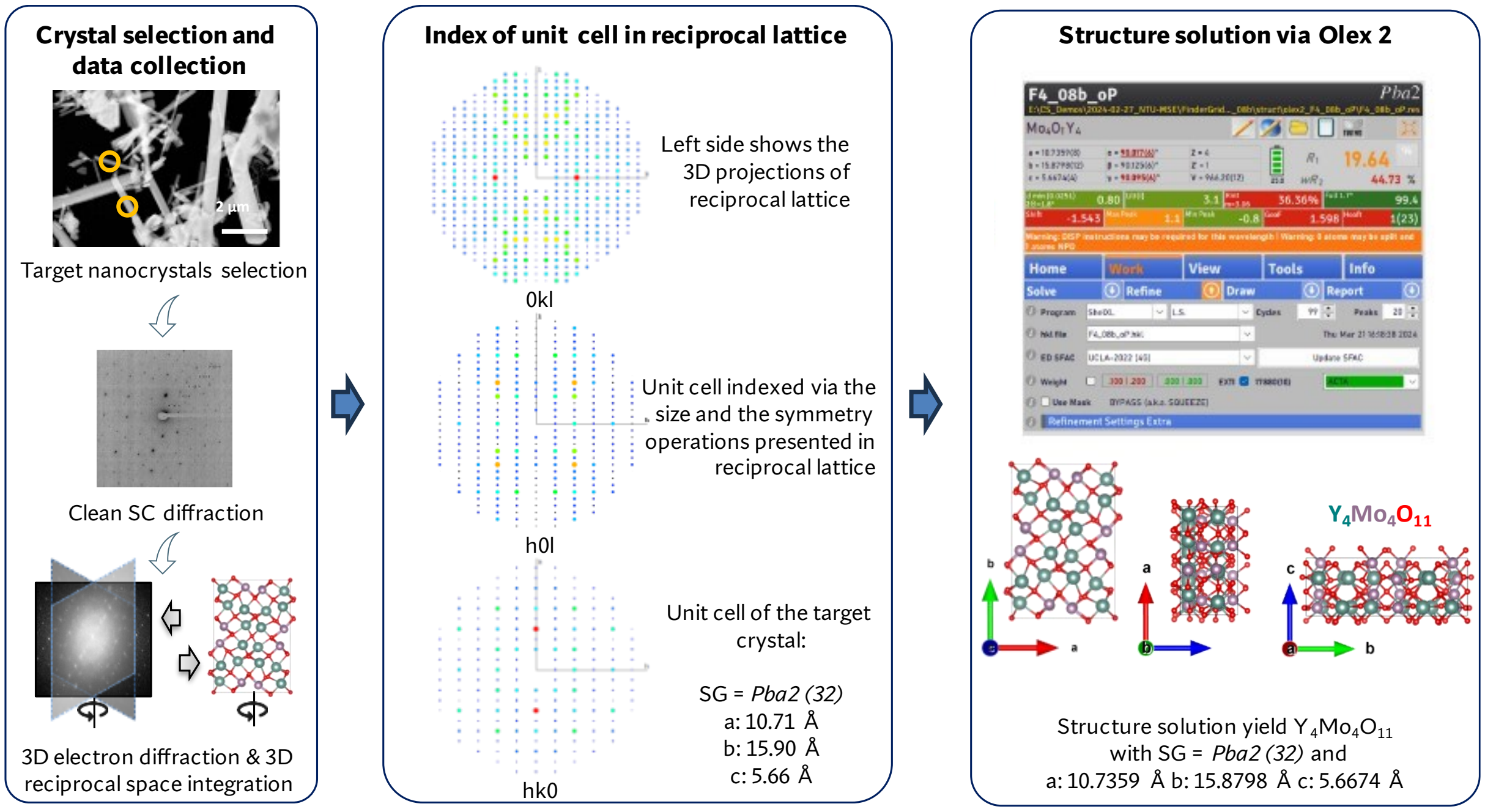

**Figure S7. Illustration of the *ab initio p*hasing from inorganic powder mixture.**

The example shows that a $YMoO_3$-like crystal exhibiting 1:1:3 elemental ratio wih new XRD diffraction peaks can be determined to be $Y_4Mo_4O_{11}$ structure.

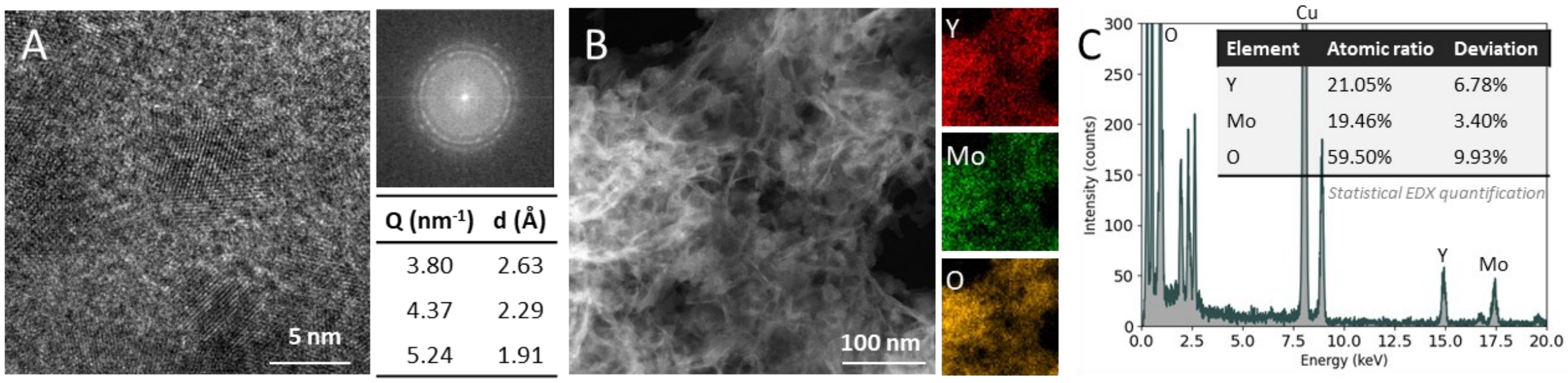

**Figure S8. Characterizations of flower-type nanocrystals with elemental ratio close to 1:1:3 produced via synthesis planned '$Na_2MoO_4$ + Mo + YClO' reaction.**

(A) HR-TEM image, FFT pattern, with list of interplanar d-spacing; (B) HADDF-STEM image with EDX maps; (C) EDX spectrum and quantification statistics. Cu signal from Cu-based TEM grid. These $YMoO_3$-like nanocrystals are 5 to 10 nm in size, with flower-like morphology in proximity to competing $Y_{2.5}MoO_6$ rods. Electron diffractograms show a polycrystalline ring pattern with d-spacing values of 2.63 Å and 2.29 Å, corresponding to the unassigned intensities observed in the refined XRD pattern (Figure S5G) at 34.0 and 39.6 degrees, respectively. Elemental quantification shows Y:Mo:O = 21.05% : 19.46% : 59.50%, closely matching that of $YMoO_3$.

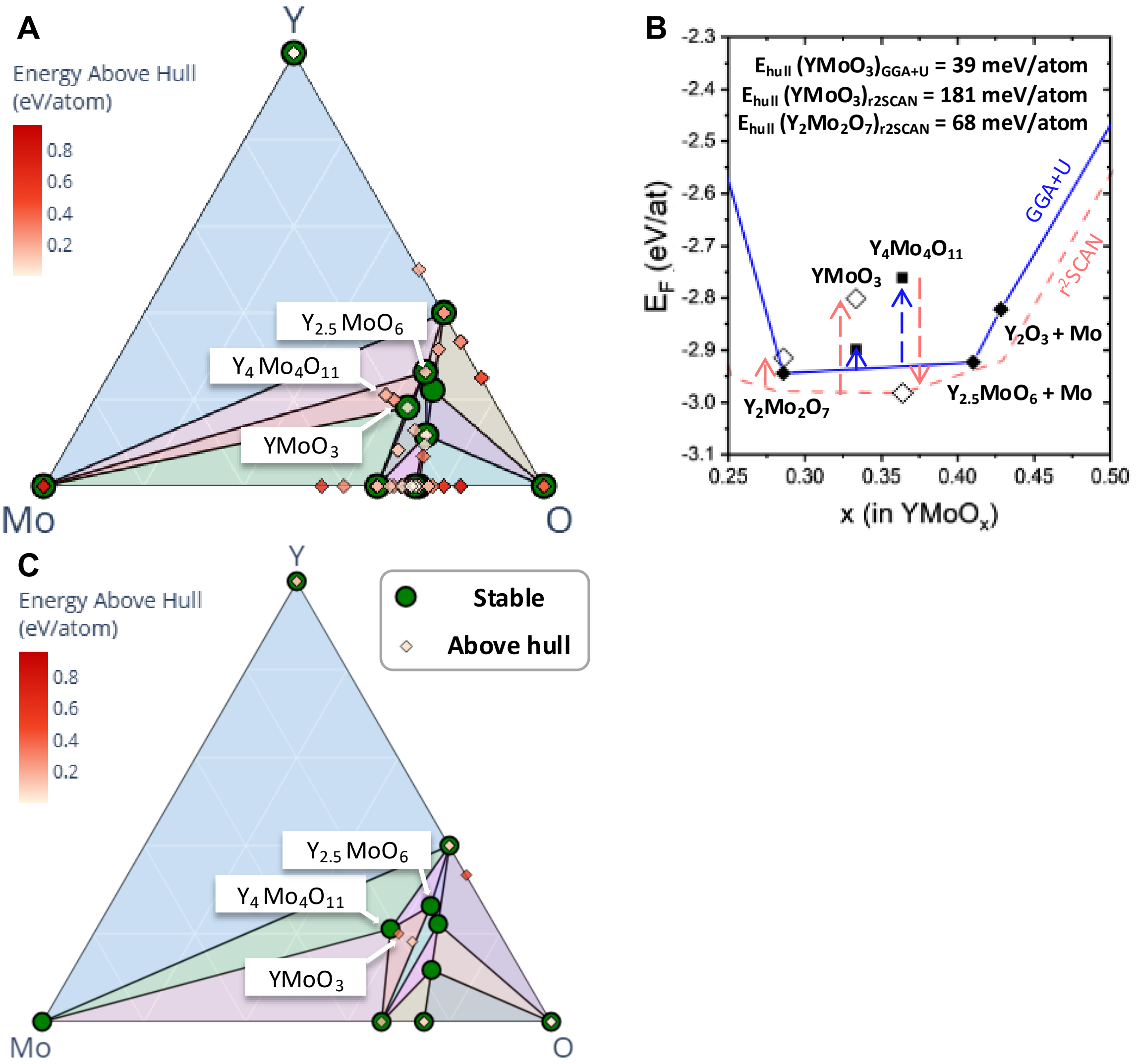

**Figure S9. DFT calculations and new convex hulls within Y-Mo-O ternary phase diagram.**

Ternary energy landscape of Y-Mo-O overlayed with $Y_{2.5}MoO_6$ phase using (A) GGA+U method and (C) metaGGA and $r^2$SCAN method. (B) Energy diagram through Y:Mo = 1:1 binary line using different DFT calculation methods (blue solid line "GGA+U", red dashed line "metaGGA and $r^2$SCAN").

**GGA+U**: $YMoO_3$ exhibits a $E_{hull}$ of 39 meV/atom, which is noticeably lower than the 170 meV/atom of $Y_4Mo_4O_{11}$, though the less stable $Y_4Mo_4O_{11}$ persists during experimentation. **metaGGA and $r^2$SCAN**: Both $Y_{2.5}MoO_6$ and $Y_4Mo_4O_{11}$ are stable on the hull surface, while a highly stable $Y_2Mo_2O_7$ lies off the hull. $YMoO_3$ exhibits a of $E_{hull}$ = 181 meV/atom, indicating that synthesizing stoichiometric, ordered $YMoO_3$ is highly challenging.

Note: the meta-GGA and r2SCAN methods offer statistically higher accuracy than GGA+U. However, at the time of this study, data using these methods remained limited in major thermochemical databases, hindering rapid computational screening.

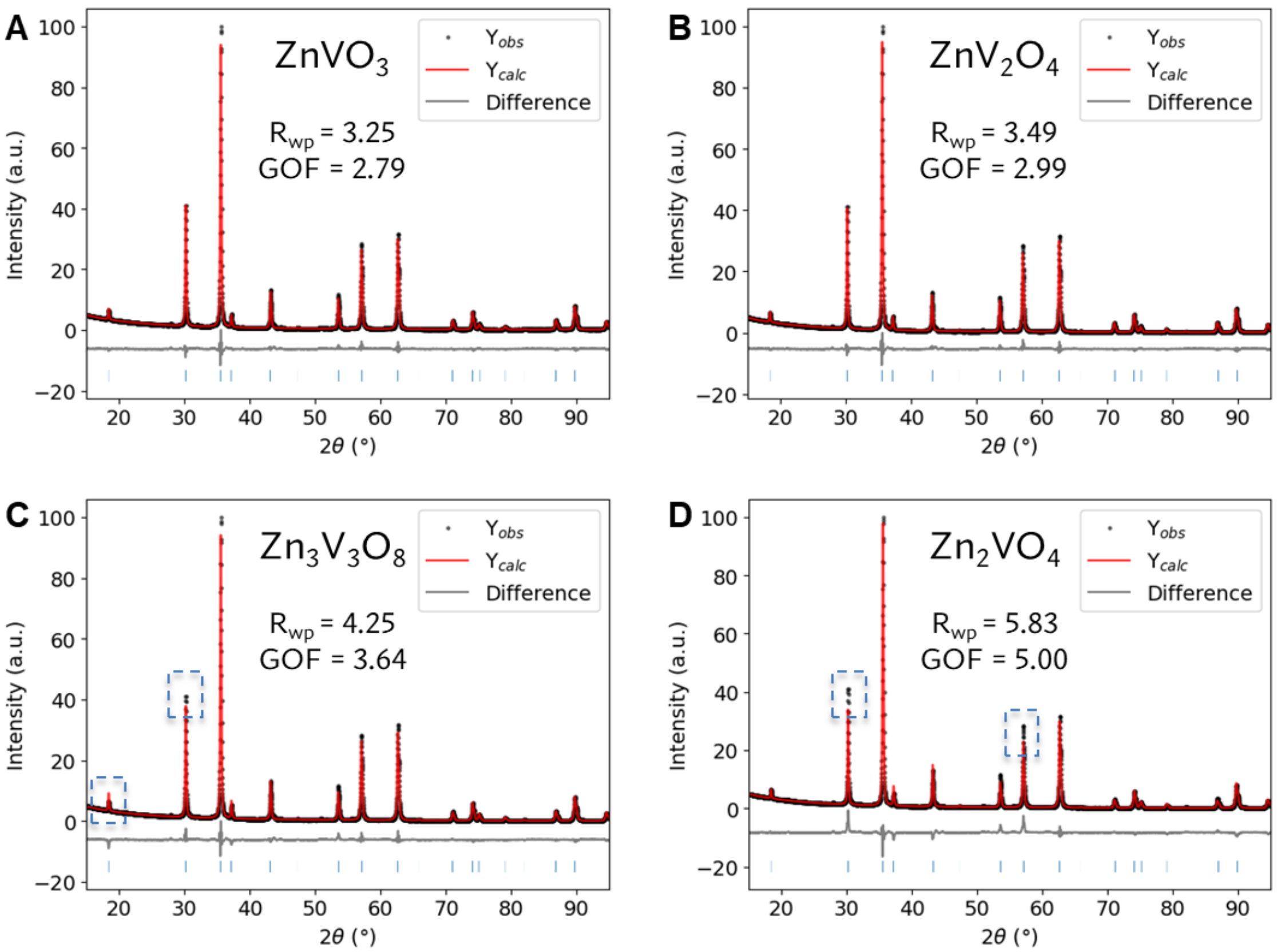

**Figure S10. Rietveld refinement of the XRD pattern of as-synthesized ZnVO3.**
(A) metal-deficient $ZnVO_3$, (B)$ZnV_2O_4$, (C) $Zn_3V_3O_8$, (D) $Zn_2VO_4$. The blue dashed boxes indicate the main source of refinement deviations.

**Table S2.** Milestone works for ML-based structure prediction and experimental realization.

| Model | Reference | Predicted | Made |
|---|---|---|---|
| GNoME (Google) | Amil Merchant *et al, Nature,* 2023 | 2.2m structures | 0 |
| A-Lab (LBNL) | Nathan J. Szymanski *et al, Nature,* 2023 | 58 structures | 41 structures (40 requiring further validation) |
| CAMD (TRI) | Joseph H. Montoya *et al, Chem Sci,* 2024 | 12 structures | 0 |
| CGNF (SNU) | Jidon Jang *et al, Matter,* 2024 | Fe-Cu-V-O phase diagram with potentially new phase | 1 (exist already) |
| MatterGen (Microsoft) | Claudio Zeni *et al, Nature,* 2025 | 8192 structures | 1 structure |
| DTMA (this work) | N.A. | 2 compositions | 1 composition |

**Table S3.** List of 7 synthesized candidate within the top 10 non-ICSD tagged candidates. (A candidate is considered synthesized if a polymorph is experimentally achieved)

| Chemical formula | Original reference |
|---|---|
| $ScFeO_3$ | *J. Am. Chem. Soc.* **136**, 15291–15299 (2014) |
| $CoVO_3$ | *Journal of Solid State Chemistry* **1**, 138–142 (1970) |
| $NiVO_3$ | *Journal of Solid State Chemistry* **2**, 521–524 (1970) |
| $CrWO_3$ | *Acta chem. scand* **8**, 932–936 (1954) |
| $ScVO_3$ | *J. Am. Chem. Soc.* **133**, 8552–8563 (2011) |
| $CrVO_3$ | *Journal of Solid State Chemistry* **144**, 392–397 (1999) |
| $AgRuO_3$ | *Chemistry A European J* **23**, 4680–4686 (2017) |

**Table S4.** Thermogravimetric analysis (TGA).

(A-B) $VO_2$ + ZnO in both inert and air environment, with corresponding theoretical weight change calculated as percentage ratio in (C).

| **(A) $VO_2$ + ZnO in Air** | | | | | |
|---|---|---|---|---|---|
| **Temperature (℃)** | **RT** | **375** | **900** | | |
| **Weight (%)** | 100.64 | 100.00 | 104.50 | | |
| **Interpretation** | Precursor | ($H_2O$ loss) | Oxidization to $Zn_2V_2O_7$ | | |
| **(B) $VO_2$ + ZnO in $N_2$** | | | | | |
| **Temperature (℃)** | **RT** | **175** | **375** | **750** | **900** |
| **Weight (%)** | 99.73 | 100.06 | 99.53 | 101.43 | 101.64 |
| **Interpretation** | Precursor | Close to $ZnVO_3$ | Close to $ZnVO_3$ | Close to $ZnVO_3$, O-rich | Close to $ZnVO_3$, O-rich |
| **(C) Theoretical weight reference** | **$VO_2$ + Zn** | **$(Zn_3V_3O_8)/3$** | **$ZnVO_3$** | **$(Zn_2V_2O_7)/2$** | |
| **Molecular weight** | 164.32 | 159.00 | 164.33 | 172.33 | |
| **Weight (%)** | 100 | 97 | 100 | 105 | |

Note: The weight change during TGA correspond to the gain of solid-state oxygen in the reaction ensemble, apart from the initial $H_2O$ release due to moisture. (Experimental 100 wt% is calibrated at 100 ℃.)

**Table S5.** STEM-EDX quantification statistics.

$YMoO_3$ synthesized via ‘$Na_2CO_3$ + $YCl_3$ + $MoO_2$’ pathway, collected from (A) JOEL 2100F, (B) ARM300F, (C) ‘$Na_2MoO_4$ + Mo + YClO’ from ARM300F, and (D) spinel $ZnVO_3$ from JOEL 2100F.

| **(A) $Y_4Mo_4O_{11}$ via ‘$Na_2CO_3$ + $YCl_3$ + $MoO_2$’ with JOEL 2100F & Oxford detector** | **O** | **Y** | **Mo** |
|---:|---|---|---|
| **S1** | 70.37% | 14.81% | 14.83% |
| **S2** | 59.21% | 21.11% | 19.68% |
| **S3** | 58.01% | 21.60% | 20.39% |
| **S4_Y2.5MoO6** | 64.15% | 25.69% | 10.15% |
| **S5** | 56.72% | 22.52% | 20.76% |
| **S6** | 51.54% | 26.27% | 22.19% |
| **S7** | 61.60% | 19.20% | 19.20% |
| **S8** | 71.61% | 12.67% | 15.73% |
| **S9_Y2.5MoO6** | 62.14% | 26.95% | 10.91% |
| **Averaged ratio** | 61.74% | 19.46% | 18.81% |
| **Standard deviation** | 6.11% | 4.12% | 1.99% |

| **(B) $Y_4Mo_4O_{11}$ via ‘$Na_2CO_3$ + $YCl_3$ + $MoO_2$’ with JOEL ARM300F & JOEL detector** | **O** | **Y** | **Mo** |
|---:|---|---|---|
| **S1** | 63.81% | 19.92% | 16.28% |
| **S2** | 57.99% | 21.81% | 20.20% |
| **S3** | 69.68% | 15.04% | 15.27% |
| **S4** | 55.10% | 23.37% | 21.53% |
| **S5** | 55.73% | 23.45% | 20.82% |
| **Averaged ratio** | 60.46% | 20.72% | 18.82% |
| **Standard deviation** | 5.54% | 3.12% | 2.54% |

| **(C) $YMoO_3$-like crystal via ‘$Na_2MoO_4$ + Mo + YClO’ with JOEL ARM300F & JOEL detector** | **O** | **Y** | **Mo** |
|---:|---|---|---|
| **S1** | 69.19% | 13.38% | 17.43% |
| **S2** | 64.02% | 17.81% | 18.17% |
| **S3** | 69.86% | 14.29% | 15.85% |
| **S4** | 54.84% | 26.24% | 18.93% |
| **S5** | 58.20% | 21.96% | 19.84% |
| **S6** | 40.88% | 32.59% | 26.53% |
| **Averaged ratio** | 59.50% | 21.05% | 19.46% |
| **Standard deviation** | 54.84% | 26.24% | 18.93% |

| **(D) Spinel crystal with structure deviation to $Zn_3V_3O_8$ JOEL 2100F & Oxford detector** | **O** | **Zn** | **V** |
|---:|---|---|---|
| **S1** | 60.02% | 20.17% | 19.81% |
| **S2** | 58.83% | 21.01% | 20.16% |
| **S3** | 52.23% | 24.66% | 23.11% |
| **S4** | 70.05% | 16.4% | 13.55% |
| **Averaged ratio** | 60.28% | 20.56% | 19.16% |
| **Standard deviation** | 6.37% | 2.94% | 3.48% |

Note: The percentages are converted to atomic ratio.

**Table S6.** Statistics of site disorder refinement for $ZnVO_3$ metal deficient spinel structure.

| | Multi. | S1 | S2 | S3 | S4 | S5 | avg. | std. | no_atom |
|---|---|---|---|---|---|---|---|---|---|
| **Zn1** | 16 | 21.12% | 21.25% | 20.65% | 20.57% | 18.29% | 20.38% | 1.08% | 0.82 |
| **Zn2** | 8 | 96.74% | 97.13% | 96.58% | 96.61% | 94.91% | 96.39% | 0.77% | 1.93 |
| **V** | 16 | 69.49% | 69.81% | 68.94% | 68.87% | 65.75% | 68.57% | 1.45% | 2.74 |
| **O** | 32 | 98.76% | 98.87% | 99.18% | 100.00% | 100.30% | 99.42% | 0.62% | 7.95 |

Note: The refinement was conducted via TOPAS v6. Five individual samples were prepared and refined in the same procedures. The refinement was conducted on the basis of parent $Zn_3V_3O_8$ with relaxed site occupancies. A equal molar constraint on Zn:V atomic ratio was added considering the elemental ratio is close to 1:1.

**Table S7.** Representative cation-deficient spinel structures with $ABO_3$ or $A_2O_3$ stoichiometry.

| Formula | Tetrahedral site (8a) | Octahedral site (16d) | Lattice constant (Å) | Lattice constant of parent phase (Å) | Unit cell expansion | ICSD |
|---|---|---|---|---|---|---|
| **$Al_2O_3$** | Al 100% | Al 83.3% | 7.9420 | Nil | Nil | 38876 |
| **$Fe_{2.689}O_4$** | Fe 92.1% | Fe 88.4% | 8.4679 | 8.3941 | 0.88% | 134312 |
| **$AlVO_3$** | Al 66.7% | Al 33.3%<br>V 66.7%<br>Total 100% | 8.4220 | 8.1928 | 2.80% | 49645 |
| **$CdSnO_3$** | Cd 81.3% | Cd 26.0%<br>Sn 66.7%<br>Total 92.7% | 9.1510 | 9.1743 | -0.25% | 23419 |
| **$FeCrO_3$** | Fe 97.0% | Fe 18.1%<br>Cr 66.7%<br>Total 74.8% | 8.2881 | 8.378 | -1.07% | 196149 |
| **$ZnVO_3$** | Zn 96.39% | Zn 20.4%<br>V 68.5%<br>Total 88.9% | 8.3959 | 8.4028 | -0.08% | Nil |

Note: The parent spinel structure refers to deficiency-free $AB_2O_4$ or $A_3O_4$ of the same elements. Site with cation deficiency (<95% occupancy) is underscored considering deviations from refinement.

**Table S8.** Statistical analysis of $A_2BB'O_6$ DFT dataset to understand the effect of neighboring compounds, including both ordered and partially disordered, to target candidate formation.

| | **With neighbor** | **Without neighbor** | **Total** |
|---|---|---|---|
| **Made experimentally** | 32.24% | 30.48% | 62.72% |
| **Not made experimentally** | 26.95% | 10.33% | 37.28% |

To assess the broader impact of partially disordered neighbors on the success rate of hypothetical candidates, we analyzed the synthesis efficacy of the $A_2BB'O_6$ database published by Chris Wolverton *et al* in 2024. which contains 2021 compositions, each structurally optimized to the lowest energy polymorph via DFT. We focused on low $E_{hull}$ (<10 meV/atom) candidates and examined neighboring effects (defined as within 10% composition deviation). From the table below, 62.72% of candidates were synthesized using formation energy alone (<10 meV/atom) as a criterion. For candidates without neighboring compositions, 30.47% were synthesized, with only 10.33% unsynthesized, suggesting neighbor competition may raise the success rate from 62% to 74%. Among unsynthesized candidates, 26.95% had neighboring compositions, while only 10.33% lacked neighbors, in contrast to synthesized candidates, where only half had neighbors. This difference disappears when considering only ordered neighboring compounds. Although we did not evaluate neighboring effects in energetic terms due to the complexity of including partially disordered structures, the data show that for $A_2BB'O_6$ oxides, neighboring partially disordered structures posed a clear influence on synthesizability, beyond formation energy considerations.

Neighbour compound is defined as within 10% of composition deviation to the target hypothetical candidate. A candidate is considered made experimentally if the candidate composition exist within ICSD or COD database. (A subset of 397 hypothetical candidates were used for this analysis focusing on low formation energy, <10 meV/atom above the convex hull. Additionally, hypothetical candidates within a blank quaternary phase diagram were not considered to avoid potentially chemically incompatible elemental combinations.)